\documentclass[
	aps,
	prd,
	reprint,
	groupedaddress,
	showpacs,
	nofootinbib,
	nobibnotes,
	amsmath,
	amssymb,
	]{revtex4-1}

\usepackage{epsfig}
\usepackage{graphicx}
\usepackage{url}
\usepackage{siunitx}

\usepackage[linktocpage]{hyperref}

\begin{document}

\title{Polyspectra searches for sharp oscillatory features in cosmic microwave sky data}

\author{J.R.~Fergusson}
\author{H.F.~Gruetjen}
\email{hfg22@cam.ac.uk}
\author{E.P.S.~Shellard}
\author{B.~Wallisch}
\email{b.wallisch@damtp.cam.ac.uk}
\affiliation{Centre for Theoretical Cosmology, DAMTP, University of Cambridge, Cambridge CB3 0WA, United Kingdom}

\date{\today}

\begin{abstract}
Despite numerous efforts, the search for oscillatory signatures in primordial spectra has not produced any convincing evidence for feature models to date. We undertake a thorough search for signatures of sharp features in the WMAP9 power spectrum and bispectrum as well as in the Planck power spectrum. For the first time, we carry out searches in both the power spectrum and bispectrum simultaneously, employing well-defined look-elsewhere statistics to assess significances in a rigorous manner. Developing efficient methods to scan power spectrum likelihoods for oscillatory features, we present results for the phenomenological bare sine and cosine modulations, allowing validation against existing Planck Likelihood surveys, as well as templates that include the correct sharp feature scaling. In particular, we study degeneracies between feature and cosmological parameters. For frequencies beyond the scale set by the acoustic peaks, the dependencies are realised through uninteresting adjustments of the comoving distance to last scattering. Hence, it is sufficient to keep cosmological parameters fixed and employ Gaussian approximations to the likelihood as a function of the feature model amplitude. In cases where results can be compared to the literature, our method shows excellent agreement. We supplement results from the Planck Likelihood with an analysis based on the Planck SMICA component separation map that, working on the assumption that the component separation algorithm is reliable, allows for the inclusion of a larger sky fraction. In principle, this allows us to place the most stringent constraints to date on the amplitudes of feature models in the temperature power spectrum. Invoking the WMAP bispectrum, we perform a combined power spectrum and bispectrum survey. We use and slightly generalise statistics developed in previous work to reliably judge the significance of large feature model amplitude estimates. We conclude that our results are entirely consistent with a featureless realisation of a Gaussian cosmic microwave background.
\end{abstract}

\pacs{98.80.Cq, 98.80.-k, 98.80.Es}

\maketitle
\tableofcontents

\section{Introduction}
\label{sec:Intro}
In recent decades, significant advances have been made in our understanding of the early Universe. The inflationary paradigm has emerged as the best explanation of how our Universe began predicting it to be flat, isotropic and homogeneous with an approximately scale invariant power spectrum of primordial fluctuations. Apart from the tensor to scalar ratio $r$, which is constrained to be less than $r<0.11$ (95\% limit) \cite{Ade:2013CosmoParam}, there is traditionally only one quantity, the spectral index $n_s$ parametrising deviations from scale invariance of the primordial power spectrum, that can be used to distinguish between differing inflationary models. Despite the observational evidence that $n_s \approx 0.96$ \cite{Ade:2013CosmoParam}, constituting a deviation from scale invariance at the four-sigma level, this constraint still leaves a plethora of viable candidates.

An exciting possibility that could provide further insight into the physics driving inflation is the presence of an oscillatory scale dependence of the primordial spectra. A well-studied scenario causing such oscillatory features are violations of the slow-roll conditions during the era of horizon exit which do not spoil inflation (cf.\ e.g.\ the review article~\cite{Chen:PrimNonGaussianities}). These can either arise due to sharp features in the slow-roll parameters (including the speed of sound) \cite{Chen:LargeNonGaussanities,Dvorkin:GSR, Miranda:WarpFeatures, Achucarro:SoundFeatures, Bartolo:EFTfeatures, Adshead:NonGaussianity}, an oscillatory component in their evolution \cite{Chen:FoldedResonant,Chen:Generation, Flauger:2009MonoInf, Flauger:Resonant} or a combination of both (cf.\ e.g.\ Refs.~\cite{Chen:StandardClock,Chen:ModelsStandardClock}). In particular, it has been shown that feature models can generate possibly observable non-Gaussianity with characteristic bispectrum shapes. This allows to look for signatures in higher-order correlation functions which can lower the threshold for a detection. It has been shown that the modulations to the power spectrum and the bispectrum are closely linked and typically oscillate with the same underlying frequency \cite{Chen:PrimNonGaussianities, Chen:FoldedResonant,Arroja:LargeStrong, Martin:ScalarBispectrum, Adshead:NonGaussianity, Achucarro:CorrelatingFeatures, Palma:UntanglingFeatures, Bartolo:EFTfeatures, Gong:CorrelatingCorrelation}.

Inspired by these ideas many searches for feature models have been undertaken in WMAP and Planck cosmic microwave background (CMB) data. Most of these focused on the power spectrum either targeting specific models, e.g.\ Refs.~\cite{Martin:OscillationsWMAP,Pahud:OscInflaton,Covi:FeaturesWMAP3,Meerburg:2011WMAP7Constraints,Peiris:2013ConstMonoInf,Flauger:2009MonoInf,Meerburg:2013SearchOscP1,Meerburg:2013SearchOscP2,Meerburg:2014SearchOsc, Achucarro:SearchCsFeatures, Ade:2013Inflation, Benetti:UpdatingConstraints, Miranda:StepsPlanck,Hu:FeaturesCMB+LSS}, or using model-independent approaches, e.g.\ Refs.~\cite{Hamann:FeaturesFrequentist, Nicholson:reconstruct, Nicholson:reconstruct2,TocchiniValentini:features,Verde:ParaReconstruction,Dvorkin:BandlimitedFeatures}. A search for oscillatory signals in the bispectrum was first undertaken in WMAP data \cite{Fergusson:Bispectrum2010} and more recently using the Planck data \cite{Ade:2013NonGaussianity}. So far, none of these searches have produced convincing evidence for the existence of features in the primordial spectra.

In this work, we will focus on the signatures of sharp features using templates that will be thoroughly discussed in Sec.~\ref{sec:featuremodels} and App.~\ref{app:sharpfeatures}. We do not only work with phenomenological sine and cosine modulations to the spectra, but also investigate whether the correct $k$-dependent scalings predicted by the theory of sharp features impact the results.

The primary goals of this work are fourfold. First, we develop efficient pipelines to scan power spectrum likelihoods for the presence of sharp features. In the process, we study degeneracies between cosmological and feature model parameters and argue that, except for very low frequencies, it is sufficient to keep cosmological parameters fixed to their best-fit values in a power spectrum survey. Secondly, we construct a likelihood based on the Planck Spectral Matching Independent Component Analysis (SMICA) \cite{Ade:2013CompSep}. While this allows for a comparison and cross-validation of the two methods, the main advantage is that the SMICA map in principle provides us with a larger accessible sky fraction. Working on the assumption that the SMICA cleaning algorithm is reliable, we can extend the included sky fraction and, thus, significantly lower the error bars on feature models. This provides clues as to whether or not large signals observed in the Planck Likelihood could be candidates for actual features. 

Thirdly, we combine the power spectrum results with WMAP bispectrum results in the spirit of a combined survey proposed in Ref.~\cite{Fergusson:psbsfeatures}. As will be discussed in detail below, finding large results at the same frequencies could provide us with further evidence for a feature. Finally, we employ and slightly generalise the statistics developed in Ref.~\cite{Fergusson:psbsfeatures} to rigorously assess the significance of the findings in the individual and combined surveys throughout this work. 

The paper is organised as follows. We start in Sec.~\ref{sec:featuremodels} by introducing and motivating the power spectrum and bispectrum templates we study. Further details are provided in App.~\ref{app:sharpfeatures}. We go on to describe and validate our pipelines in Sec.~\ref{sec:Methods}. In particular, we study degeneracies between the cosmological and the additional feature model parameters in the power spectrum in Sec.~\ref{subsec:Methods-PowerSpectrum}. While Sec.~\ref{subsubsec:PlanckLike} is focused on the Planck Likelihood, our power spectrum analysis of the Planck SMICA component separation map is detailed in Sec.~\ref{subsubsec:FastQuadEst}. Section~\ref{subsec:Methods-Bispectrum} then provides an outline of the bispectrum pipeline employed on the WMAP data.

Our main results are presented in Sec.~\ref{sec:results}. Sections~\ref{subsec:Results-PowerSpectrum} and \ref{subsec:Results-Bispectrum} discuss the results of individual Planck power spectrum and WMAP bispectrum surveys. In particular, we compare the power spectrum results obtained from the Planck Likelihood with those obtained from the SMICA map for different sky fractions to investigate whether the inclusion of more data reinforces the observed signals. Section \ref{subsec:Results-combined} goes on to combine the power spectrum results from the WMAP Likelihood as well as the Planck Likelihood and SMICA map with the WMAP bispectrum. Throughout Sec.~\ref{sec:results} we employ the statistics developed in Ref.~\cite{Fergusson:psbsfeatures} to rigorously judge the significance of our findings and discuss to what extent the data is consistent with a featureless Gaussian CMB. To do this we slightly generalise the work in Ref.~\cite{Fergusson:psbsfeatures} as detailed in App.~\ref{app:statistics}. Finally, we summarise our results and conclude in Sec.~\ref{sec:summconc}.

\section{Feature models: oscillating polyspectra}
\label{sec:featuremodels}
As in a previous publication \cite{Fergusson:psbsfeatures}, we continue to study a linearly-spaced template with the power spectrum $P_{\mathcal{R}}(k)$ and the bispectrum shape $S(k_1,k_2,k_3)$ given by
\begin{align}
\label{eq:barePS}
\frac{\Delta P_{\mathcal{R}}}{P_{\mathcal{R},0}}(k)=&A_P\sin{(2\omega k+\phi_P)}\,,\\
\nonumber
S(k_1,k_2,k_3)=&\frac{(k_1 k_2 k_3)^2}{\Delta^4_{\mathcal{R}}(k_{*})}  B(k_1,k_2,k_3)\\
\label{eq:bareBS}
=&A_B\sin(\omega K + \phi_B)\,,
\end{align}
where $K=k_1+k_2+k_3$, $P_{\mathcal{R},0}(k)$ is the power spectrum in the absence of any feature, $\Delta^2_{\mathcal{R}}(k)=k^3/(2\pi^2)P_{\mathcal{R},0}(k)$ is the dimensionless power spectrum and $k_{*}$ is a fiducial momentum scale. We will refer to this template as PS1, Eq.~\eqref{eq:barePS}, and BS1, Eq.~\eqref{eq:bareBS}, respectively, to distinguish them from the modified templates, PS2 and BS2, that will be introduced below. Note that for a given feature the same frequency $\omega$ appears in both the modulation to the power spectrum and the oscillatory running of the bispectrum. The frequency $\omega$ is a dimensionful quantity with units of Mpc, but, for brevity, we suppress units when quoting frequencies throughout this work. The phases $\phi_P$ and $\phi_B$ and especially the amplitudes $A_P$ and $A_B$ are typically model-dependent so that this template has five parameters.

Such oscillatory spectra are well-motivated theoretically. While other models are also known to produce oscillations%
\footnote{See for example Refs.~\cite{Danielsson:TransPlanckPhysics,Danielsson:VacuumChoice} for a discussion of non-standard vacuum choices that produce sinusoidal modulations to the power spectrum.}%
, a well-explored example is the appearance of sharp features in the slow-roll parameters or the speed of sound during inflation. A rigorous treatment in the case of the power spectrum involves the generalised slow-roll (GSR) framework (cf.\ e.g.\ Ref.~\cite{Dvorkin:GSR}) while the bispectrum is studied using the in-in formalism (cf.\ e.g.\ Ref.~\cite{Chen:PrimNonGaussianities}) with possible GSR corrections to the mode functions \cite{Adshead:FastCompBispec}. 

In this work, we are mostly interested in the behaviour of the oscillatory spectra for $\omega k\gg1$. The reason for this is that for extended oscillations, the contribution to the overall S/N (signal to noise) in the CMB from low multipoles is small, so that nearly all the S/N comes from $l>\mathcal{O}(10^2)$ corresponding to $k>\mathcal{O}(10^{-2})$. Low frequencies $\omega\lesssim 140$ are strongly degenerate with cosmological parameters as shown explicitly in Sec.~\ref{subsubsec:PlanckLike} and we will be mainly interested in large frequencies with $\omega\gg10^2$. Thus, for these models most of the S/N generically comes from regions with $\omega k\gg1$ so that it is justified to scan for these models based on their behaviour for $\omega k\gg1$.

Rather than rigorously calculating polyspectra employing the GSR method and the in-in formalism, we provide a simplified discussion in App.~\ref{app:sharpfeatures} that allows us to extract the leading-order behaviour for $\omega k\gg1$ and motivates our templates. In the case of the power spectrum, simple solution matching across a sharp feature, as is done in App.~\ref{subapp:sharpfeaturesPS}, shows that one generally expects a feature in single-field inflation to generate linearly-spaced oscillations where the leading-order behaviour for $\omega k\gg1$ is given by a component $\sim\cos(2\omega k)$ and a suppressed component $\sim\sin(2\omega k)/(\omega k)$. This is consistent with rigorous results from the GSR approximation in the literature \cite{Miranda:WarpFeatures}. The relative magnitude of these terms is determined by the jumps in the slow-roll parameters at the location of the feature. This statement is general in the sense that it also applies to inflation with non-standard kinetic terms where sharp features in the speed of sound can occur. In fact, features in $\epsilon$ and $c_s$ are (nearly) degenerate at the level of the power spectrum%
\footnote{In the context of the GSR framework this degeneracy between features in the inflaton potential and in the speed of sound has been demonstrated for example in Ref.~\cite{Miranda:WarpFeatures}. The degeneracy is only broken by minor differences at small $k$ that are observationally nearly irrelevant due to the poor S/N at low multipoles and do not contribute to our discussion of the $\omega k\gg1$ behaviour.}. 

The case of the bispectrum is briefly discussed in App.~\ref{subapp:sharpfeaturesBS}. Extracting the leading-order behaviour with the in-in formalism shows that the scaling of the bispectrum for a sharp feature in the slow-roll parameters is given by a component $\sim (\omega K)^2\cos(\omega K)$ and a suppressed component $\sim (\omega K)\sin(\omega K)$ in single-field inflation. While steps in the speed of sound give rise to oscillatory signals in the bispectrum as well, they typically produce a different shape, hence breaking the near degeneracy found in the power spectrum. In this work, we focus on the former case.

We see that even though the templates PS1 and BS1, Eqs.~\eqref{eq:barePS} and~\eqref{eq:bareBS}, have the correct oscillatory behaviour that we encounter in the phenomenology of sharp features, they do not correctly capture the scalings of all the components. Hence, we also search for modulations that include the correct scalings using the templates
\begin{align}
\label{eq:modPS}
\frac{\Delta P_{\mathcal{R}}}{P_{\mathcal{R},0}}=&A_P\left(\cos{\phi_P}\frac{f_P(\omega)}{\omega k}\sin{(2\omega k)+\sin{\phi_P}\cos{(2\omega k)}}\right)\,,\\
\nonumber
S=&A_B\left(\cos{\phi_B}f_B(\omega)(\omega K)\sin(\omega K)\right.\\
\label{eq:modBS}
&\qquad\qquad\qquad\left.+\sin{\phi_B}(\omega K)^2\cos(\omega K)\right)\,,
\end{align}
where $f_P(\omega)$ and $f_B(\omega)$ are functions chosen to give equal S/N to the sine and the cosine components and the angles $\phi_P, \phi_B\in [0,\pi)$ parametrise the relative magnitudes. These templates will be referred to as PS2, Eq.~\eqref{eq:modPS}, and BS2, Eq.~\eqref{eq:modBS}, respectively.

We emphasise that while these templates should be reliable for $\omega k\gg 1$, the behaviour of the solutions for $\omega k\lesssim 1$ is much more complicated and requires a careful treatment. We argued previously that this region contributes very little S/N. However, in the case of the template PS2, the $1/(\omega k)$ suppression can assign a larger fraction of the overall S/N to this region. To avoid this problem we will restrict our analysis for the template PS2 to multipoles $l\ge 50$, enforcing that for $\omega\gg100$ only $k$ with $\omega k\ge 1$ contribute to feature model amplitudes. This solution clearly comes with a loss in S/N for those feature models that have most of their support in the region $l\le 50$. These models would benefit from a more rigorous treatment. We will discuss this point in more detail in Sec.~\ref{subsubsec:PS2survey}.  

We conclude this section with a few remarks on the validity of the sharp feature limit (cf.\ App.~\ref{app:sharpfeatures}). As has been discussed in Refs.~\cite{Adshead:Bounds, Cannone:PertUnitarity}, strictly speaking, the sharp feature limit is not under perturbative control. This can be seen naively by considering the ratio of the quadratic and cubic Lagrangian that should satisfy $\mathcal{L}_3/\mathcal{L}_2\ll1$. The couplings in the cubic Lagrangian diverge as the sharp feature limit is taken and violate this bound which indicates that the theory becomes strongly coupled. Hence, features that can be studied within the framework of perturbation theory can only have a small, but finite width. The finite width typically manifests itself as an exponentially decaying envelope multiplying the feature templates. Effects on wavenumbers $k$ which were deep inside the horizon at the time of the feature are suppressed. For the frequency range we are studying it is sensible to assume that very sharp, but still perturbative, features produce a signature in the power spectrum that is largely unaffected by the envelope in the signal-dominated region of Planck. Thus, these types of features should be well described by the templates PS1 and PS2 without an envelope. We are using WMAP data with $l_{\text{max}}=600$ to study the bispectrum. In this case it is also sensible to assume that very sharp, but still perturbative, features are captured by the templates BS1 and BS2 without taking the envelope into account.

\section{Methods}
\label{sec:Methods}
\subsection{Power spectrum: a dual pipeline}
\label{subsec:Methods-PowerSpectrum}
The analysis of the power spectrum is performed employing a dual pipeline based on the Planck 2013 data release. Our search for feature models in the power spectrum uses both the Planck power spectrum Likelihood as well as a pseudo-$C_l$ (PCL) likelihood based on the Planck component separation maps \cite{Ade:2013CompSep}. While the former incorporates a more rigorous modelling of experimental effects such as noise anisotropies and beam uncertainties, the latter allows, for example, to easily change the included sky fraction and use larger parts of the CMB sky that, assuming successful component separation, should be clean of foregrounds. We provide an outline of the two approaches in Secs.~\ref{subsubsec:PlanckLike} and~\ref{subsubsec:FastQuadEst}, respectively.

As previously pointed out and confirmed in this work, the likelihood function has many local maxima associated with different frequencies $\omega$. This is a serious obstacle for a standard Markov Chain Monte Carlo (MCMC) analysis as the likelihood is difficult to explore. Therefore, we introduce a grid in the frequency $\omega$ as has been done in previous works \cite{Meerburg:2011WMAP7Constraints, Meerburg:2013SearchOscP1, Meerburg:2013SearchOscP2, Fergusson:psbsfeatures}. The spacings were taken to be $\Delta\omega=10$ unless stated otherwise. We also introduce a grid in the phase $\phi$ with $\Delta\phi=0.1\pi$. However, as will be further discussed in Sec.~\ref{subsubsec:phasecomment}, adopting the methods discussed below it usually suffices to perform the analysis only for the pure sine component, $\phi=0$, and the pure cosine component, $\phi=\pi/2$. All other phases are straightforwardly related to these two cases.

For any given frequency $\omega$ and phase $\phi$ we only vary the amplitude $A_P$ and find the best fit while keeping all cosmological, foreground and nuisance parameters fixed and set to their best-fit values without the presence of any feature. We will argue and verify in Sec.~\ref{subsubsec:PlanckLike} that this is entirely sufficient for a search for linearly-spaced feature models with frequencies $\omega\gg 100$ which are beyond the oscillation patterns imprinted on the CMB due to the acoustic oscillations. 

\subsubsection{Planck Likelihood}
\label{subsubsec:PlanckLike}
\paragraph{Fast extraction of feature model amplitudes from the Planck Likelihood.}
\label{paragraph:FastExtraction}
Unsurprisingly, keeping the six cosmological $\Lambda$CDM parameters $A_s$, $n_s$, $\Omega_b h^2$, $\Omega_c h^2$, $\theta_A$ and $\tau$ as well as the foreground and nuisance parameters fixed, the Planck Likelihood is very nearly Gaussian%
\footnote{While the high-$l$ part of the likelihood should be Gaussian as pointed out below, there can be deviations due to the non-Gaussian nature of the low-$l$ likelihood. Furthermore, non-linear corrections to the modulations of the $C_l$ from lensing might cause very slight deviations from a perfect Gaussian shape of the likelihood as a function of $A_P$.} %
in the feature model amplitude $A_P$ for any given $\omega$ and $\phi$. We verified this empirically by plotting the likelihood, but it can also be understood in a straightforward fashion by noting that the high-$l$ log-likelihood, which dominates the S/N, is based on the fiducial Gaussian approximation \cite{Efstathiou:MythsandTruths, HamimecheLewis:CMBlikelihoods, Ade:2013Likelihood} given by
\begin{equation}\label{eq:fidGauss}
-2\log{\mathcal{L}}=\chi^2\equiv\left(\hat{C}_{l_1}-C_{l_1}\right)\Delta_{l_1l_2}\left(\hat{C}_{l_2}-C_{l_2}\right)\,,
\end{equation}
where $\hat{C}_l$ and $\Delta_{l_1l_2}=\langle \Delta\hat{C}_{l_1}\Delta\hat{C}_{l_2}\rangle$ are the PCL estimates and covariance matrix, respectively. The covariance matrix is evaluated for a fiducial model and kept fixed, explaining the name. If we only vary the feature model amplitude, we have $C_l=C_{0,l}+A_P\delta C_l$ in linear theory and, therefore, it is evident that $\mathcal{L}$ as a function of $A_P$ is a Gaussian.

Rather than thoroughly exploring the likelihood using MCMC techniques, it is thus possible to simply calculate the best-fit amplitude $\hat{A}_P$ and the variance $\sigma^2=\langle\Delta\hat{A}_P^2\rangle$ by fitting a Gaussian through three points. In practice, we sample the likelihood on a coarse grid in the amplitude $A_P$ using a version of \textsc{CAMB} \cite{Lewis:CAMB} modified to include the additional feature degrees of freedom. In order to correctly resolve the oscillations over the entire frequency range, we make sure that the accuracy settings are chosen appropriately%
\footnote{For increasing frequency, we adjust the \textsc{CAMB} accuracy parameters `accuracy\_boost' and `l\_sample\_boost' to enforce a denser wavenumber sampling, a decrease in integration step sizes and a denser sampling in $l$ when interpolating the $C_l$ \cite{Howlett:CAMB}. We explicitly checked that our settings are sufficient to resolve oscillations at a given $\omega$ by ensuring that our results are unaffected by a further increase in accuracy parameters. A detailed study of accuracy settings in the context of feature searches can be found in Ref.~\cite{Meerburg:2011WMAP7Constraints}. We generally chose to be more conservative using $(\mathrm{accuracy\_boost},\mathrm{l\_sample\_boost})=(3,30)$ for $\omega<600$, $(4,40)$ for $\omega<1000$ and $(8,50)$ for $\omega\ge 1000$. Especially at lower frequencies this is likely to be excessively accurate and further optimisation is possible.}. %
In particular, we enforce calculation of the transfer functions at each $l$ for large frequencies $\omega\ge 1000$. Then, we pick the amplitude with the lowest $\chi^2$ from this small set of samples%
\footnote{This is just a rough estimate for the actual best-fit amplitude due to the limited number of samples from the likelihood. We emphasise that the actual best-fit amplitude is calculated by fitting a Gaussian.} %
and two points approximately a distance $\sigma$ to the left and right and calculate the corresponding Gaussian. The mean of this Gaussian is the best-fit amplitude $\hat{A}_P$ and its variance gives $\langle\Delta\hat{A}_P^2\rangle$. The whole process is very fast making scans over large frequency ranges feasible.

We also performed an MCMC exploration with \textsc{CosmoMC} \cite{Lewis:CosmoMC} only varying $A_P$ for frequencies up to $\omega=2000$ enforcing a stringent convergence criterion of $R-1<0.01$ to exclude the possibility that non-Gaussian corrections to the likelihood have a large effect. We found excellent agreement of the mean and standard deviation of the posterior amplitude distribution with the Gaussian approximation described above in accordance with the expectation that the likelihood should be very nearly Gaussian.

To quantify our results, we assign a significance $\bar{A}_P$ to the best-fit amplitudes according to
\begin{equation}
\label{eq:barAdef}
\bar{A}_P=\frac{\hat{A}_P}{\langle\Delta\hat{A}_P^2\rangle^{\frac{1}{2}}}\,.
\end{equation}
From a Bayesian point of view this significance measures how inconsistent the posterior amplitude distribution is with $A_P=0$. However, this significance can also be interpreted from a frequentist point of view. Under the fiducial Gaussian approximation, the maximum-likelihood estimate $\hat{A}_P$ is normally distributed with mean $\langle\hat{A}_P\rangle=0$ under the null hypothesis ($A_P=0$) with the same variance $\langle\Delta\hat{A}_P^2\rangle$. Hence, Eq.~\eqref{eq:barAdef} is also the frequentist significance corresponding to the $p$-value of measuring an amplitude at least as big as $\hat{A}_P$. Finally, we note for later reference that for the fiducial Gaussian approximation mentioned above, the likelihood improvement, which is often quoted, is simply given by
\begin{equation}
-2\Delta\log{\mathcal{L}}=-\bar{A}_P^2\,.
\end{equation}

\paragraph{Comparison with a full MCMC analysis.}
\label{paragraph:FullMCMC}
The analysis outlined above does not explore the possibility of further likelihood improvements coming from varying the cosmological parameters. Hence, one might be worried that the method does not identify big likelihood improvements or, equivalently, significant feature model amplitudes with large $\bar{A}_P$ reliably. The purpose of this section is to show that for frequencies larger than the scale of acoustic oscillations, $\omega\gg 100$, varying the amplitude alone is sufficient to correctly determine the significant feature model signals and the maxima in likelihood improvement.

To do this, we compare results obtained from the Planck Likelihood by first fixing all cosmological, foreground and nuisance parameters to their best-fit values and then just varying the feature model amplitude as above to an analysis, where the six cosmological and 14 foreground and nuisance parameters%
\footnote{We note that the other foreground and nuisance parameters incorporated in the Planck Likelihood are analytically marginalised over as in Ref.~\cite{Ade:2013CosmoParam}.} %
are varied simultaneously with the amplitude $A_P$. For the latter, we assume the same priors on the cosmological, foreground and nuisance parameters as the Planck Collaboration (see Tables~1 and~4 in Ref.~\cite{Ade:2013CosmoParam}) and a flat prior with $-0.3\leq A_P\leq 0.3$ on the feature amplitude. We run eight \textsc{CosmoMC} chains until they satisfied the convergence criterion $R-1<0.02$ in the least-converged parameter with the actual convergence being closer to 0.01 and below for most cases. Additionally, we also checked that the results are unaffected by a more stringent convergence criterion.

\begin{figure}[t]
\includegraphics[width=\columnwidth]{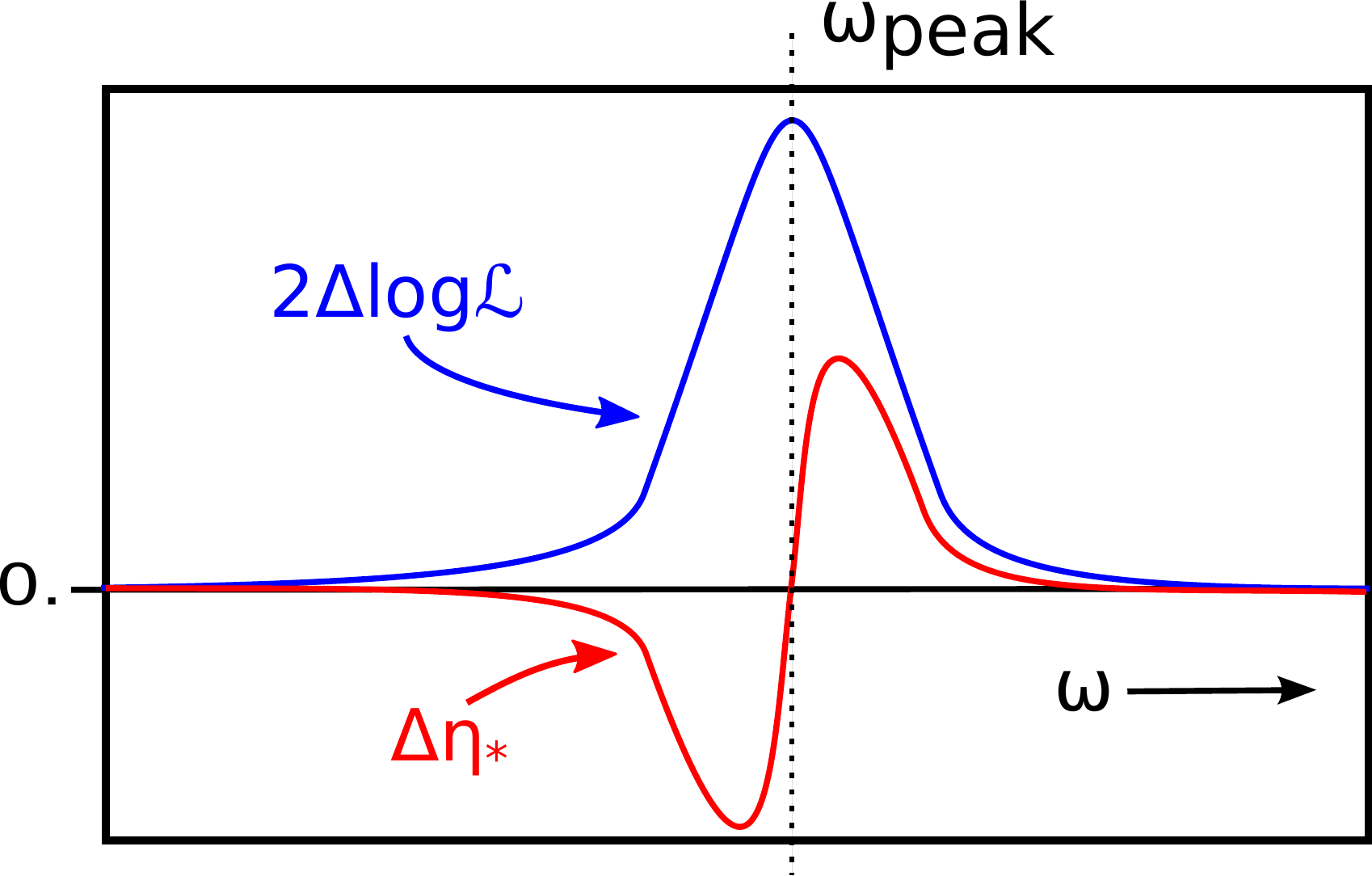}
\caption{An idealised sketch of the behaviour of the comoving distance to LS $\eta_{*}$ in the vicinity of a peak in the feature model likelihood. The blue curve represents the likelihood improvement as a function of $\omega$ measured when only varying the amplitude and keeping all cosmological parameters fixed to their best-fit values. The red curve shows the behaviour of $\Delta\eta_{*}=\eta_{*}-\eta_{*,0}$ attempting to tune the resulting $l$-space frequency $\omega_l$ to match $\omega_{l,\text{peak}}$.}
\label{fig:LSdistsketch}
\end{figure}

Let us first formulate some expectations to interpret the results. As already discussed in Ref.~\cite{Fergusson:psbsfeatures}, we expect strong degeneracies between feature models and the cosmological parameters below the scale set by the sound horizon at last scattering (LS), $\omega\approx140$. The oscillations due to features effectively alter the acoustic peak structure which can mimic the effects of changing other parameters. These degeneracies are expected to largely disappear for $\omega>140$. However, as has been mentioned elsewhere (see e.g.\ Ref.~\cite{Meerburg:2013SearchOscP1}), even for high-frequency oscillations there still is a further subtle effect that can lead to dependencies between the feature model amplitude and the cosmological parameters. For a feature model with primordial frequency $\omega$ in $k$-space, the frequency $\omega_l$ in $l$-space observed in the power spectrum is approximately given by
\begin{equation}
\omega_l\sim\frac{2\omega}{\eta_{*}}\,,
\end{equation}
where $\eta_{*}$ is the comoving distance to LS. The latter depends on the expansion history of the Universe and thus on the $\Lambda$CDM parameters. For a given point on the grid of feature models, the frequency $\omega$ is fixed. However, if we decide to vary the cosmological parameters, we effectively allow some freedom in the frequency that is ultimately compared to the data. If there is a large signal at some $\omega_{l,\text{peak}}$ in the data, we see a peak at $\omega_{\text{peak}}\sim \eta_{*,0}\omega_{l,\text{peak}}/2$ in the computations where the cosmological parameters are kept fixed. Here, $\eta_{*,0}$ is the best-fit value of $\eta_{*}$ without allowing for feature models that we used in these runs. For this particular value $\omega=\omega_{\text{peak}}$ we expect that allowing the cosmological parameters to vary will not have any impact on the best fit. It should still have the same feature model parameters and we should still find a best fit with $\eta_{*,BF}=\eta_{*,0}$. However, if the frequency is slightly lower than $\omega_{\text{peak}}$, we expect that the best-fit cosmological parameters get shifted such that $\eta_{*}$ decreases in an attempt to arrange $\omega_l\sim 2\omega/\eta_{*}\sim\omega_{l,\text{peak}}$. Similarly, if $\omega$ is slightly larger than $\omega_{\text{peak}}$, we expect $\eta_{*}$ to increase in order to produce a better fit to the feature in the data. These expectations are summarised in Fig.~\ref{fig:LSdistsketch}%
where the behaviour of $\Delta\eta_{*}=\eta_{*}-\eta_{*,0}$ around a peak in the likelihood improvement is shown schematically.

\begin{figure*}[t]
\includegraphics[width=1.5\columnwidth]{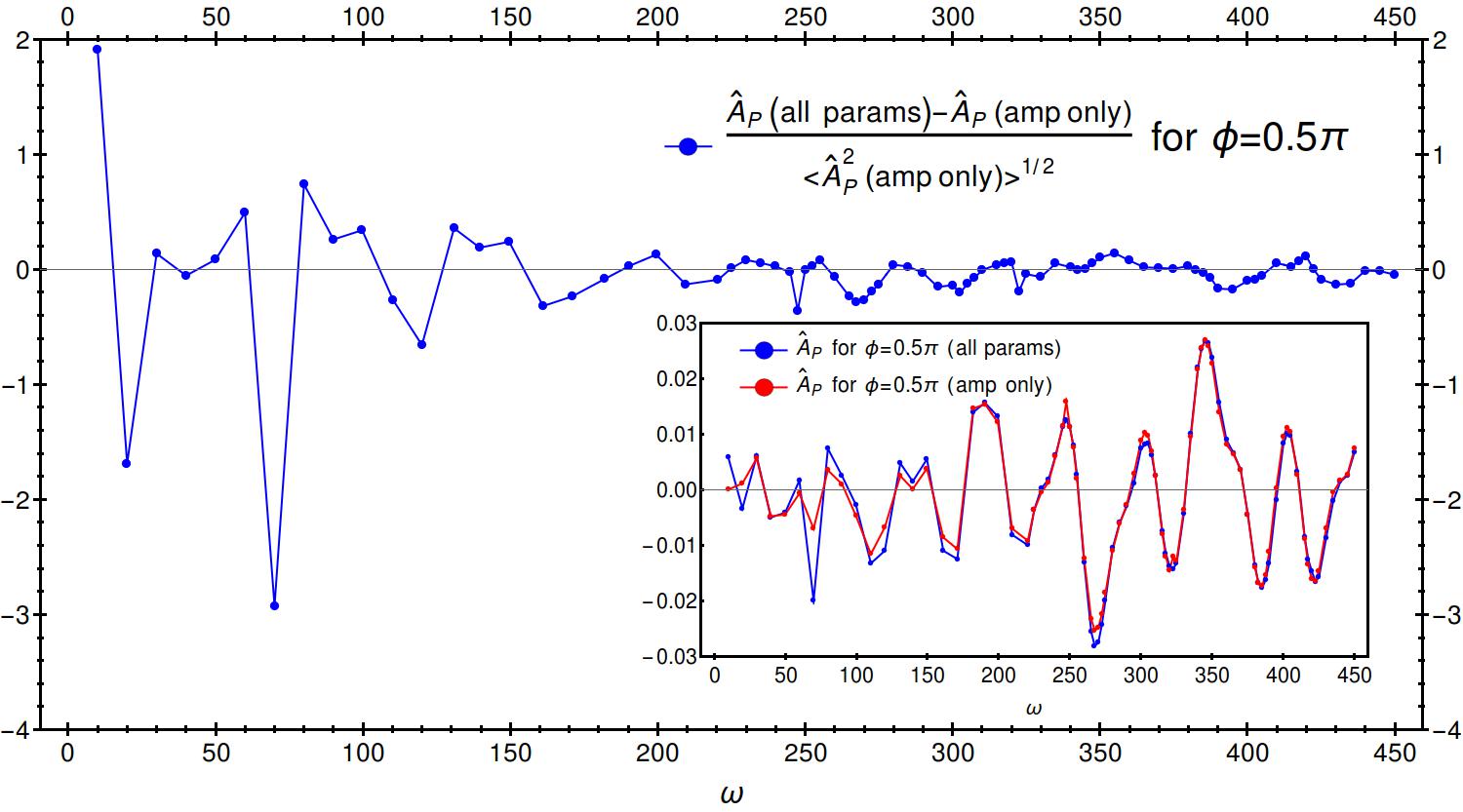}
\caption{Comparison of the oscillation amplitudes obtained from an MCMC analysis of the Planck Likelihood sampling over all cosmological, foreground and nuisance parameters (all params) and the fast method where all parameters are kept fixed and the likelihood is assumed to be Gaussian in the amplitude $A_P$ (amp only). In the former case, we take $\hat{A}_P$ to be the mean of the posterior amplitude distribution as we find this to be numerically more stable than the best fit. In the latter case, we use the best-fit amplitudes obtained from the Gaussian approximation. We checked that this gives almost exactly the same answer as the mean of the corresponding posterior distribution due to the near perfect Gaussianity of the likelihood in this case (see discussion in the main text). For each frequency the results are displayed for $\phi=0.5\pi$. The main plot shows the difference between the amplitudes normalised by the standard deviation as calculated from the Gaussian approximation. A direct comparison of the amplitudes is shown as an inset. Good agreement is found for $\omega\gtrsim200$.}
\label{fig:AmplitudeComparison}
\end{figure*}

With this heuristic picture in mind we proceed to study results extracted from the Planck Likelihood. In this section, we decrease the frequency spacing to $\Delta\omega=5$ for $\omega\in[220,450]$ (even smaller values are chosen in the vicinity of the likelihood peaks) in order to better resolve the peaks in the amplitude estimates and expected jumps in $\Delta\eta_{*}$. Figure~\ref{fig:AmplitudeComparison}%
compares the amplitudes obtained with and without varying the cosmological, foreground and nuisance parameters. The amplitude estimates $\hat{A}_P$ are plotted for $\phi=0.5\pi$ where the feature template reduces to a bare cosine, $\Delta P_\mathcal{R} \sim \cos(2\omega k)$. For $\omega\gtrsim200$ this shows good agreement between the values obtained from the two approaches. We emphasize that this is not only true for the slice shown, but also for all other values of the oscillation phase $\phi$. Evidently, any remaining degeneracies have little effect on the measured amplitudes for large frequencies.

\begin{figure*}[t]
\includegraphics[width=1.8\columnwidth]{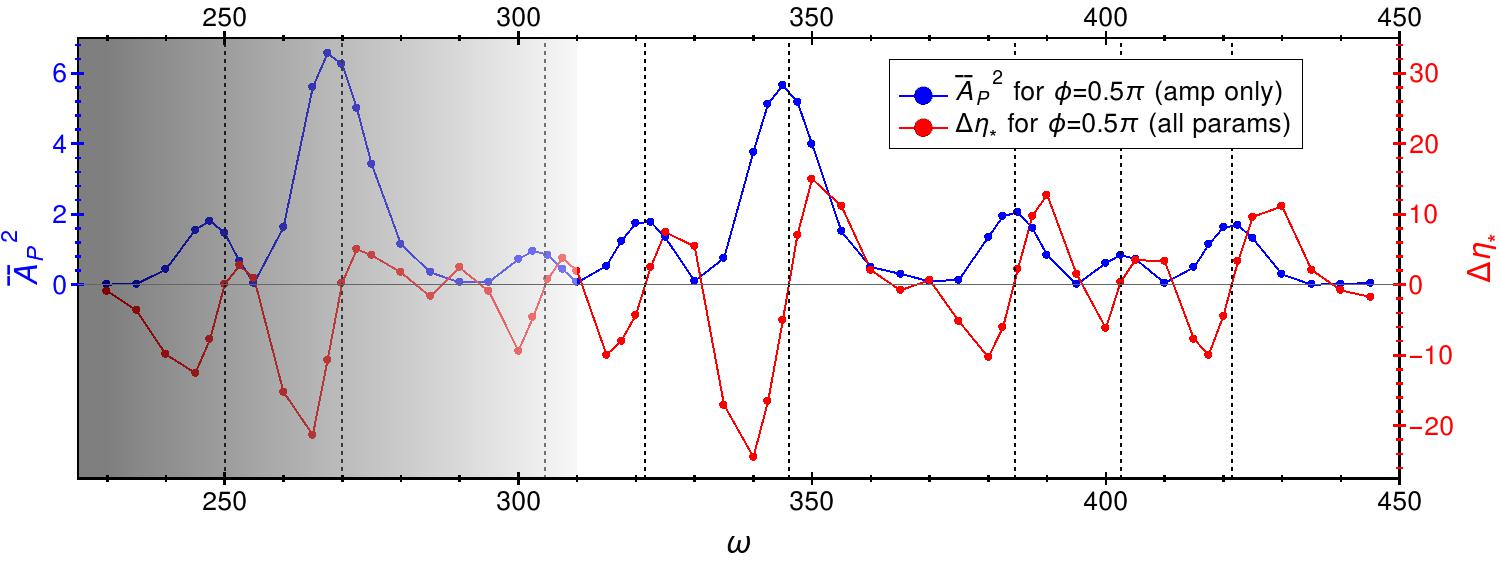}
\caption{The behaviour of the comoving distance to LS, $\eta_{*}$, as a function of frequency $\omega$ (red). $\eta_{*}$ is obtained by varying all cosmological, foreground and nuisance parameters as well as the amplitude $A_P$. We plot $\Delta\eta_{*}=\eta_{*}-\eta_{*,0}$ where $\eta_{*,0}$ is the best-fit value setting $A_P=0$. Also plotted is the likelihood improvement $\bar{A}_P^2$ obtained by keeping all parameters fixed and employing a Gaussian approximation to the likelihood as discussed in the main text (blue). For each frequency the result is plotted for $\phi=0.5\pi$. The shaded area indicates where one expects further residual degeneracies with cosmological parameters and the idealised picture presented in Fig.~\ref{fig:LSdistsketch} does not fully apply. For $\omega\gtrsim300$ one can clearly identify the predicted behaviour of $\eta_{*}$ in the vicinity of the likelihood peaks. Dashed vertical lines are drawn at the approximate locations of peak frequencies $\omega_{\text{peak}}$ as a guide.}
\label{fig:SoundHorizonComparsion}
\end{figure*}

However, the degeneracy with $\eta_{*}$ discussed above can indeed be observed at higher $\omega$. Figure~\ref{fig:SoundHorizonComparsion}%
shows the likelihood improvements for $\omega$ in the range $[230,445]$ obtained from just varying the amplitude together with the values of $\eta$ obtained from runs where all parameters are varied. The qualitative behaviour shown in Fig.~\ref{fig:LSdistsketch} is clearly recognisable at the likelihood peaks for $\omega\gtrsim300$. For smaller frequencies this effect can still be observed, but is less prominent due to more complicated degeneracies in this region not captured by this simple model. The presence of these residual low-$\omega$ degeneracies is also suggested by the small differences in the amplitude measurements near the peaks in the region $200\lesssim\omega\lesssim300$ in Fig.~\ref{fig:AmplitudeComparison}.

Summing up, for $\omega\gg140$ a search for features where only the amplitude is varied is sufficient and any remaining degeneracies with cosmological parameters are small. Our results suggest that any further effects can be explained by an adjustment of the comoving distance to LS. Such an effect does not lead to bigger peaks in the likelihood improvement. It only allows frequencies close to a peak to benefit slightly, but does not change the likelihood improvement measured for the actual peak frequency.

\subsubsection{Fast quadratic estimate based on the SMICA map}
\label{subsubsec:FastQuadEst}
To search for feature models in the Planck SMICA component separation map \cite{Ade:2013CompSep} we construct a simple PCL likelihood as in previous work \cite{Fergusson:psbsfeatures}. Using the SMICA map has the advantage that we can extend the included sky fraction compared to the Planck Likelihood \cite{Ade:2013Likelihood, Ade:2013CosmoParam} allowing in principle for more stringent tests of feature models%
\footnote{Whether the results are reliable depends on the performance of the foreground cleaning scheme in regions of higher contamination. We will discuss this point in more detail below.}. 

As has become standard, we use a PCL likelihood based on cross-correlators to analyse the power spectrum \cite{Larson:WMAP7powerspectra, Ade:2013Likelihood, Ade:2013CosmoParam}. The Planck component separation analysis published two maps for each foreground cleaning method as part of the 2013 data release. A map of the full CMB temperature sky (which is supposed to be clean of foregrounds) that we will refer to as $T(\hat{n})$ and a half-ring-half-difference map that we refer to as $N(\hat{n})$. The latter is obtained by running the foreground cleaning pipeline on the data from the first and second half of each stable pointing period and taking half the difference of the results. It can be thought of as an estimate of the noise in the final temperature map \cite{Ade:2013CompSep}.

The data enters the likelihood through the power spectrum estimates
\begin{equation}
\hat{C}_l^X=\hat{C}_l^T-\hat{C}_l^N\,,
\end{equation}
where $\hat{C}_l^T$ and $\hat{C}_l^N$ are the power spectra of the $T$ and $N$ maps and we use the standard PCL estimates obtained via
\begin{equation}
\hat{C}_{l_1}=(M^{-1})_{l_1l_2}\tilde{C}_{l_2}\,,\quad \tilde{C}_l=\frac{1}{2 l+1}\sum\limits_{m}\vert a_{lm}\vert^2\,,
\end{equation}
with $M_{l_1l_2}$ the standard PCL coupling matrix for a given mask \cite{Hivon:MASTER}. The subscript $X$ indicates that we think of this power spectrum effectively as a cross spectrum. The rationale behind this is the following: If we define the maps
\begin{equation}
T_1(\hat{n})=\frac{T(\hat{n})+N(\hat{n})}{2},\quad T_2(\hat{n})=\frac{T(\hat{n})-N(\hat{n})}{2}
\end{equation}
and think of the $T$ map as the half-ring-half-sum map, then the half-ring (HR) maps $T_1$ and $T_2$ correspond to the foreground cleaned maps obtained from the first and second half of each stable pointing period. The cross-correlator between these two maps is then given by
\begin{align}\nonumber
\tilde{C}_l^{T1\times T2}&=\frac{\sum_m (a^1_{lm})^*a^2_{lm}}{2l+1}\\
&=\frac{\sum_m \left((a^T_{lm})^*a^T_{lm}-(a^N_{lm})^*a^N_{lm}\right)}{2l+1}\equiv\tilde{C}_l^X\,.
\end{align}
This argument is obviously only entirely correct if the foreground cleaning procedures are exactly linear in all input maps, which is not the case for all foreground cleaning schemes. However, for the purpose of this analysis we will assume this is a good approximation and treat $\hat{C}_l^X$ as a cross-correlator that should then also carry no noise bias%
\footnote{There is another caveat here that there can be correlations between the noise in the first and second half of the pointing periods. Such correlations spoil the independence of the noise realisation in the two HR maps and would thus lead to a noise bias in the cross-correlator. This minor issue leads to small underestimates of the noise in the HR maps and was discussed in Ref.~\cite{Ade:2013CompSep}. We will ignore this here.}.

As in Ref.~\cite{Fergusson:psbsfeatures} we apodise the masks by approximate convolution with a Gaussian beam of FWHM $0.5^{\circ}$ to minimise leakage using the procedure outlined in Ref.~\cite{Gruetjen:inpainting}. We approximate the PCL log-likelihood with the fiducial Gaussian approximation \cite{Efstathiou:MythsandTruths, HamimecheLewis:CMBlikelihoods, Ade:2013Likelihood} introduced in Eq.~\eqref{eq:fidGauss}. As the fiducial model we simply use a smoothed version of the power spectrum $\hat{C}_l^X$ together with a smoothed version of the power spectrum $\hat{C}_l^N$ to model the noise contribution to the covariance%
\footnote{The fiducial model and the noise spectrum only enter the covariance approximation and should roughly correspond to the true power spectrum underlying the data and the actual noise power spectrum. To obtain good approximations, we simply use the PCL estimates of these spectra given by $\hat{C}_l^X$ and $\hat{C}_l^N$. As we are only including multipoles with $l\ge 50$ the scatter in these estimates is not very significant. However, to further reduce this scatter, we smooth the PCL spectra $l(l+1)\hat{C}_l^X$ and $\hat{C}_l^N$ over a width $\Delta l\approx 20$ by convolution with a Gaussian.}. %
We employ the analytic approximations from Ref.~\cite{Efstathiou:MythsandTruths} to calculate the covariance matrices. These approximations assume an approximately constant power spectrum. To account for small leakage effects we correct for slight underestimates of the variance using an improved analytic approximation \cite{Gruetjen:inpainting}.

In order to calculate the covariance matrices we have to assume a noise model for the HR maps. In the analysis below we just model the noise as isotropic and Gaussian with a power spectrum given by a smoothed version of $2\,\hat{C}_l^N$. While the noise is clearly not isotropic in reality%
\footnote{Anisotropy in the noise arises at the very least due to the highly anisotropic scanning strategy of Planck. To a lesser extent it is also conceivable that the foreground cleaning methods introduce anisotropies in the noise patterns of the final maps.}, %
we find that this assumption does not seem to affect the feature searches presented in this work at any significant level%
\footnote{To test this, we also constructed a likelihood assuming an anisotropic noise pattern based on the average hit counts of the 143 and 217 GHz maps. While the $\chi^2$ values shift slightly, the significances with which various feature models are detected seem to be largely unaffected. Therefore, we stick to a simple isotropic model.}. %
Due to significant deviations from a Gaussian distribution at low $l$ the fiducial Gaussian approximation is not reliable in this region. Hence, we only consider the multipole range $50\le l\le2000$. As we are mainly interested in extended oscillations, the loss in S/N from discarding the low multipoles is negligible. Naively, assuming $l_{\text{max}}\sim 2000$, the region $2\le l\le 50$ should contribute only a fraction of $\mathcal{O}(10^{-3})$ to the total S/N. However, this estimate does not take into account the specific shape of the oscillatory feature model templates that are increasingly damped in amplitude due to the convolution with the transfer functions and lensing. Furthermore, it is well known that there is a dip in the power spectrum at $20\lesssim l\lesssim 30$ \cite{Ade:2013CosmoParam, Ade:2013Likelihood} that may have some weight. To quantify this, we explicitly checked the influence of the low-$l$ likelihood on the results for the PS1 template in the Planck Likelihood analysis and found shifts in the amplitudes of at most $\sim 0.3$ sigma and usually far below. This is a rather small effect although larger than the naive S/N consideration suggests, which is likely due to the anomaly mentioned previously.

In linear theory, especially ignoring lensing, the observed CMB power spectrum given the six $\Lambda$CDM parameters $p_i$ and a feature model is given by
\begin{equation}
C_l(p_i, A)=C_l^{\Lambda\text{CDM}}(p_i)+A_P\,\delta C_l(p_i,\omega, \phi)\,.
\end{equation}
If $A_P\delta C_l$ is a small contribution to the power spectrum, which it always is in this work, the lensed power spectrum can be written in the same way by linearising the effect of lensing and defining $\delta C_l:=\partial C_l^{\text{lensed}}/\partial A_P$.
In principle, determining the maximum-likelihood estimate for the amplitude $\hat{A}_P\equiv\hat{A}_P^{\text{ML}}$ requires all parameters to be varied simultaneously. As we showed in Sec.~\ref{subsubsec:PlanckLike}, it is sufficient for the purpose of this study to set the $\Lambda$CDM parameters to their best-fit values obtained by assuming a featureless model and only vary the amplitude $A_P$ for any given $\omega$ and $\phi$. We thus compute the lensed feature model templates for each point on the grid assuming the best-fit $\Lambda$CDM cosmology employing \textsc{CAMB} \cite{Lewis:CAMB} with sufficiently high precision settings to ensure that the oscillations are accurately resolved.

As we only vary the amplitude $A_P$, the best fit can be found as a simple quadratic estimate%
\footnote{We introduce a redundant factor of 2 in the definition of the quadratic estimator here for consistency with the standard optimal power spectrum estimator and the optimal bispectrum estimator later on. This definition implies $\langle\hat{A}^2_P\rangle=2!/N_P$ in line with $\langle\hat{A}_B^2\rangle=3!/N_B$ for the bispectrum.} %
given by
\begin{align}
\label{eq:quadest}
\hat{A}_P&=\frac{2}{N_P}\delta C_{l_1}(\Delta^{-1})_{l_1l_2}\left(\hat{C}_{l_2}-C^{\Lambda\text{CDM}}_{l_2}\right)\,,\\
N_P&=2\delta C_{l_1}(\Delta^{-1})_{l_1l_2}\delta C_{l_2}
\end{align}
for any $\omega$ and $\phi$. The variance of the estimates is simply given by $\langle\hat{A}_P^2\rangle=2/N_P$ so that we can extract the normalised amplitude estimates, Eq.~\eqref{eq:barAdef}, via
\begin{equation}\label{eq:barAquadest}
\bar{A}_P=\left(\frac{2}{N_P}\right)^{\frac{1}{2}}\delta C_{l_1}(\Delta^{-1})_{l_1l_2}\left(\hat{C}_{l_2}-C^{\Lambda\text{CDM}}_{l_2}\right)\,.
\end{equation}
This leads to very quick scans over the full frequency range. The estimator is essentially the optimal quadratic estimator for the amplitude $A_P$ except for well-known and small suboptimalities due to the slightly lossy data compression in PCL power spectrum estimates (cf.\ Ref.~\cite{Gruetjen:TowardsEfficient} and references within).

\subsubsection{A comment on the phase \texorpdfstring{$\phi$}{phi}}
\label{subsubsec:phasecomment}
Even though the methods discussed above are sufficiently fast that one can simply introduce a grid in the phase $\phi$ as well and obtain significances $\bar{A}_P(\omega,\phi)$ for each point on a sufficiently dense $(\omega,\phi)$-grid, we emphasise that this is typically not necessary. This is particularly evident from Eq.~\eqref{eq:barAquadest}. For all feature models considered in this work it can be written as
\begin{equation}
\bar{A}_P(\phi)=\frac{N_P(0)^{\frac{1}{2}}}{N_P(\phi)^{\frac{1}{2}}}\cos{\phi}\bar{A}_P(0)+\frac{N_P(\pi/2)^{\frac{1}{2}}}{N_P(\phi)^{\frac{1}{2}}}\sin{\phi}\bar{A}_P(\pi/2)
\end{equation}
suppressing the frequency $\omega$. The normalisation factors can be easily calculated and the dependence on the data is only through the two estimates $\bar{A}_P( 0)$ and $\bar{A}_P(\pi/2)$. Further simplifications occur due to the fact that the sine and cosine components are in all cases very nearly uncorrelated already for moderately high frequencies, $\langle \hat{A}_P(\omega, 0)\hat{A}_P(\omega, \pi/2)\rangle\approx0$, and, furthermore, we have $\langle\hat{A}^2_P(\omega, 0)\rangle\approx\langle\hat{A}^2_P(\omega, \pi/2)\rangle$ . The latter holds automatically for the bare sine and cosine modulations of the template PS1 and is arranged through the $f_P(\omega)$ factor in the case of the template PS2. This implies $N_P(\phi)\approx N_P(0)\approx N_P(\pi/2)$ so that we arrive at the simple relation
\begin{align}\nonumber
\bar{A}_P(\omega, \phi)&=\frac{\hat{A}_P(\omega, \phi)}{\langle\hat{A}^2_P(\omega, \phi)\rangle^{\frac{1}{2}}}\\
&=\cos{\phi}\bar{A}_P(\omega, 0)+\sin{\phi}\bar{A}_P(\omega, \pi/2)\,.
\end{align}
Even though we made use of the form of the quadratic estimator, we expect that the same reasoning can be applied to the Planck Likelihood owing to the fact that it is largely based on the fiducial Gaussian approximation that gave rise to Eq.~\eqref{eq:quadest}. We checked this explicitly and for all the results presented in Sec.~\ref{sec:results} this is an extremely good approximation.

Note that with minimal modifications, this discussion also applies to the bispectrum estimator that will be discussed below. References~\cite{Munchmeyer:KSWestimator,Munchmeyer:OptimalEstimator} make use of this property in their search for oscillatory models in the bispectrum where it is sufficient to evaluate the estimator for the sine and the cosine component only.

\subsection{Bispectrum: optimal \texorpdfstring{$f_{\mathrm{NL}}$}{FNL} estimator}
\label{subsec:Methods-Bispectrum}
To constrain feature models via the bispectrum we use a modified version of the modal polynomial pipeline that was used in the 2013 Planck analysis \cite{Ade:2013CosmicStrings, Ade:2013NonGaussianity, Ade:2013IsotropyStatistics, Ade:2013ISW}. This is an implementation of the standard optimal bispectrum estimator. The optimal estimator for the bispectrum amplitude of a feature model $\hat{A}_B$, in the diagonal covariance approximation, reads
\begin{widetext}
\begin{equation}
\hat{A}_B=\frac{1}{N_B}\sum_{l_i m_i}  \frac{\mathcal{G}^{l_1 l_2 l_3}_{m_1 m_2 m_3} b_{l_1 l_2 l_3}  \left(a_{l_1 m_1} a_{l_2 m_2} a_{l_3 m_3} - 3\langle a_{l_1 m_1} a_{l_2 m_2}\rangle a_{l_3 m_3}\right)}{ C_{l_1} C_{l_2} C_{l_3}}\,,
\end{equation}
\end{widetext}
where $b$ is the theoretical bispectrum of the feature model defined by
\begin{equation}
\langle a_{l_1 m_1} a_{l_2 m_2} a_{l_3 m_3} \rangle = B^{l_1l_2l_3}_{m_1m_2m_3}=\mathcal{G}^{l_1 l_2 l_3}_{m_1 m_2 m_3} b_{l_1 l_2 l_3}\,
\end{equation}
and $\mathcal{G}$ is the Gaunt integral, which is the projection of the angular part of the primordial delta function. The normalisation of the estimator is
\begin{equation}\label{eq:norm}
N_B \equiv \sum_{l_il_i}\frac{\left(\mathcal{G}^{l_1 l_2 l_3}_{m_1 m_2 m_3} b_{l_1 l_2 l_3}\right)^2}{ C_{l_1} C_{l_2} C_{l_3} } \,.
\end{equation}
The pipeline breaks the bispectrum being constrained into a set of orthonormal separable basis bispectra which dramatically reduces computation time and allows us to constrain all frequencies, within resolution, simultaneously. The approach was first described in Ref.~\cite{Fergusson:PrimordialNonGaussianity} and a fully realised version was first implemented in Ref.~\cite{Fergusson:BispectrumEstimationI}. It was recently extended to polarisation in preparation for the next round of Planck papers in Ref.~\cite{Fergusson:Efficient}, which also included many other small advances. Here, we use the temperature only version of this pipeline on WMAP 9-yr data \cite{Bennett:WMAP9Maps} restricting ourselves to $l_{\text{max}}=600$. This reduction in range coupled with the increase in the number of basis functions from 600 to 2000 allows us to cover a frequency range six times larger than in the first Planck analysis extending up to $\omega = 1000$. For the WMAP data we use the weighted average of the V and W channels with weights \num{1.0} and \num{0.9}, respectively. The linear term and variance is computed from 500 simulations generated with the fiducial power spectrum combined with white anisotropic noise created from the coadded hit count maps. The simulations were then masked and diffusively inpainted to reduce mode coupling in the multipoles. We use a frequency grid with a stepwidth of $\Delta \omega=20$, which is sufficient for a WMAP-type survey as discussed in Sec.~\ref{subsec:Results-Bispectrum} and App.~\ref{app:statistics}.

\section{Results}
\label{sec:results}

\subsection{Power spectrum surveys}
\label{subsec:Results-PowerSpectrum}

\subsubsection{Bare sine and cosine: results for the template PS1}
\label{subsubsec:PS1survey}
\begin{figure*}
 \centering
\includegraphics[width=\textwidth]{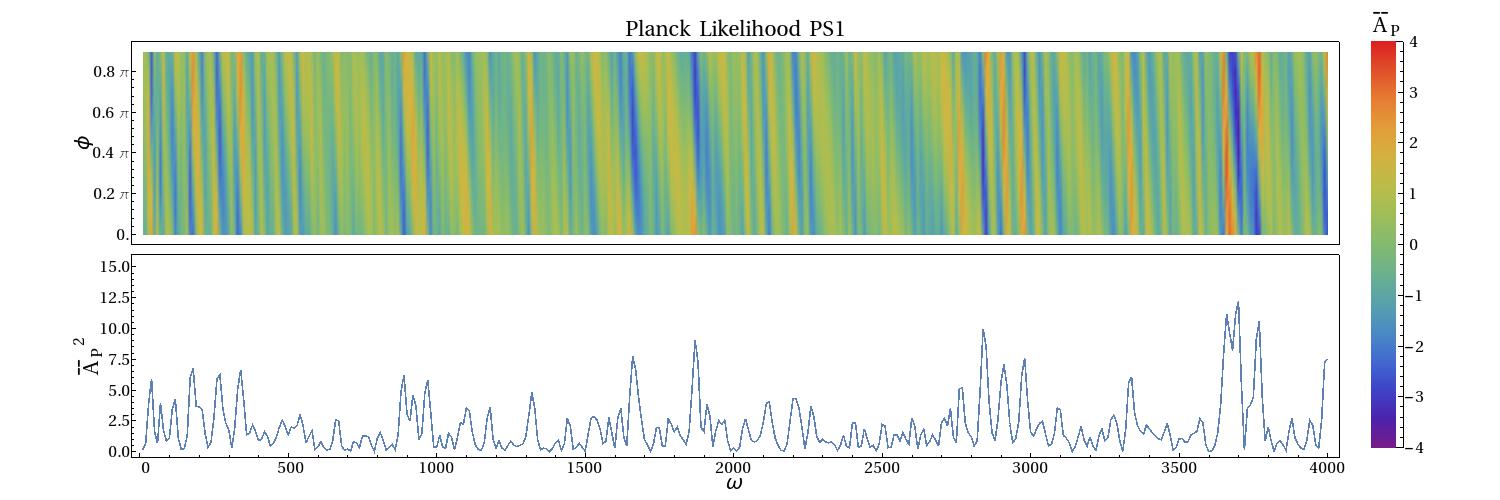}
\includegraphics[width=\textwidth]{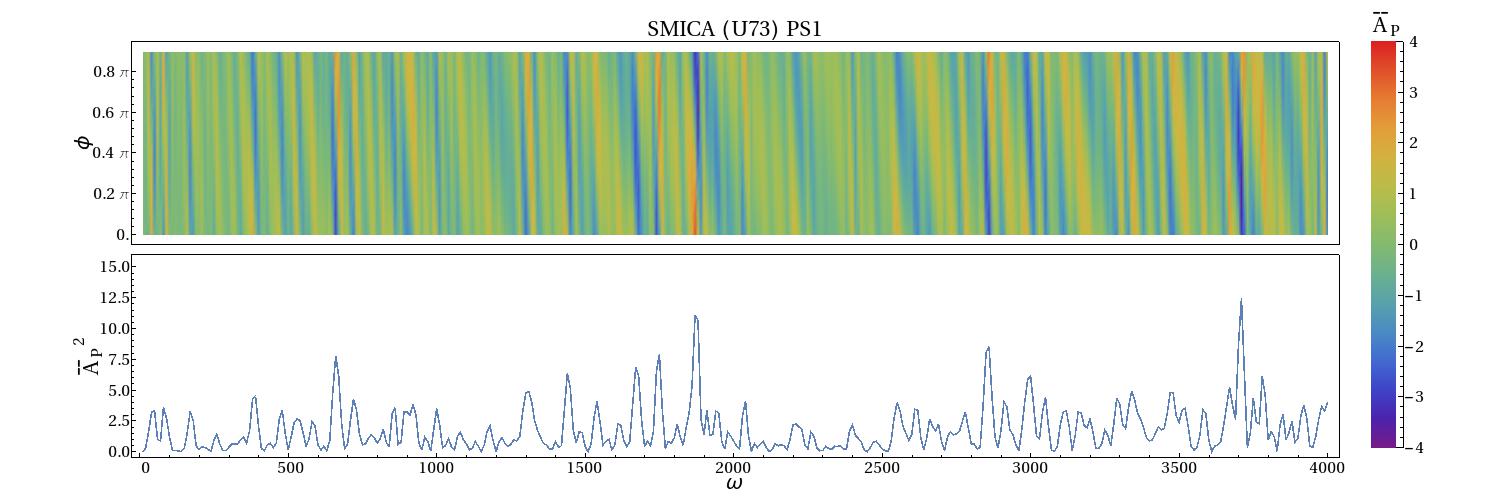}
\caption{Amplitudes $\bar{A}_P$ for the template PS1 up to $\omega=4000$ as obtained from the Planck Likelihood (top) and based on the SMICA map masked with an apodised version of the U73 mask with $f_{\text{sky}}=0.61$ (bottom). In each plot the bottom panel shows the maximum $\bar{A}_P^2$ at a given $\omega$, corresponding to the maximum likelihood improvement at that frequency.}
\label{fig:ampso4000}
\end{figure*}

We present the results for the template PS1 for both the Planck Likelihood and the SMICA map masked with the U73 mask in Fig.~\ref{fig:ampso4000}.%
Each plot consists of a top panel with a density plot of the normalised amplitude $\bar{A}_P$ as defined in Eq.~\eqref{eq:barAdef} and a bottom panel showing the maximum $\bar{A}_P^2$ found for any phase $\phi$ at a given frequency $\omega$. We remind the reader that the latter corresponds to the maximum likelihood improvement observed at that frequency.

The results show clear similarities. However, there are evidently differences. We emphasise that due to the different sky fractions included in the analysis one should not expect the results to match exactly%
\footnote{In addition, the low-$l$ likelihood is included in the Planck Likelihood search, but not in the SMICA analysis. As already discussed in Sec.~\ref{subsubsec:FastQuadEst}, this can give rise to further minor deviations in the measured amplitudes.}. %
Even if we assume that a given method exactly accounts for all systematics and noise properties of the data, inclusion of more data necessarily shifts the observed peaks. The Planck Likelihood is based on the CL49 and CL31 mask with a sky fraction of $f_{\text{sky}}=0.49$ and $f_{\text{sky}}=0.31$, respectively, so that significantly more data is included%
\footnote{One should keep in mind that sky fractions do not fully reflect the difference in the underlying datasets. The CL31 and CL49 masks are not simply larger versions of the U73 mask, but there are parts of the sky that are masked by the U73 mask, but not by the CL31 or CL49 mask. This leads to larger differences between the results than might be expected just based on the ratio of sky fractions.} %
in the SMICA analysis with $f_{\text{sky}}=0.61$. Note that this implies that, assuming the foreground cleaning procedure is reliable, tighter constraints on feature models can be obtained from the SMICA map (cf.\ Fig.~\ref{fig:variances}). We quantified how correlated the results are and arrived at the conclusion that the differences between the Planck Likelihood and SMICA results can be explained by the differences between the masks used in the two analyses.

The results for the Planck Likelihood are in very good agreement with the corresponding Fig.~3 in Ref.~\cite{Meerburg:2013SearchOscP2} where a similar search has been performed using a different method. Their approach relies on a Taylor expansion of the power spectrum in the cosmological parameters which allows for faster sampling when varying all parameters as the derivatives of the transfer functions can be precomputed \cite{Meerburg:2013SearchOscP1}. Then, a Metropolis-Hastings algorithm is used to find the best fit. By adopting sampling schemes tailored to the problem, further improvements in computational efficiency are possible \cite{Meerburg:2014SearchOsc}. Note that we argued previously in Sec.~\ref{paragraph:FullMCMC} that this is not necessary for linearly-spaced oscillations and accurate amplitude estimates as well as likelihood improvements can be extracted using the numerically very efficient and reliable methods proposed in Sec.~\ref{sec:Methods} that keep cosmological parameters fixed to their best-fit values. The striking agreement with Ref.~\cite{Meerburg:2013SearchOscP2} is a further validation of this claim.

We find the overall best fit at $\omega\sim3710$ using both our approaches with a significance of about \num{3.5} sigma corresponding to $2\Delta\log{\mathcal{L}}\approx 12$. Reference~\cite{Meerburg:2013SearchOscP2} reports the maximum likelihood improvement to be $2\Delta\log{\mathcal{L}}\approx 13$ at $\omega=3670$ (note that their frequency definition differs from ours by a factor of two), which is part of the same peak structure%
\footnote{The peak observed at this frequency in the Planck Likelihood analysis is only marginally smaller than our best fit and recorded at \num{3.3} sigma ($2\Delta\log{\mathcal{L}}\approx 11$) making it a competing maximum. In fact, the authors of Ref.~\cite{Meerburg:2013SearchOscP2} also obtained this structure with three nearby large peaks, whose likelihood improvements differ slightly leading to a different overall best-fit value.}.

The natural question that arises when studying the results is whether or not the large signals in various places present significant evidence for feature models. A generic property of feature model surveys is that one scans over a large number of effectively independent models. Each of these can fit fluctuations in the noise%
\footnote{Noise in this context refers to both the cosmic variance and experimental noise. Both cause scatter in the PCL estimates that can give rise to good feature model fits by chance.} %
by chance so that we expect large results in a large survey simply because we compared a vast number of models to the data. The look-elsewhere effect that arises for the oscillatory feature models under investigation was studied in Ref.~\cite{Fergusson:psbsfeatures}. There it was argued that the distribution of the maximum significances observed in an individual survey can be well described analytically and gives rise to a look-elsewhere-adjusted significance according to
\begin{equation}\label{eq:sigind}
S=2^{\frac{1}{2}}\text{Erf}^{-1}\left[\left(F_{\chi,\,2}\left(\bar{A}_P\right)\right)^{N_{\text{eff}}}\right]\,.
\end{equation}
Here, $N_{\text{eff}}$ quantifies the effective number of independent models and $F_{\chi,\,2}(x)$ is the cumulative distribution function (CDF) of the $\chi$-distribution with two degrees of freedom. Equivalently, one can define an effective step width in frequency $\Delta\omega_{\text{eff}}$ according to
\begin{equation}\label{eq:domeffdef}
\Delta\omega_{\text{eff}}=\frac{\omega_{\text{max}}-\omega_{\text{min}}}{N_{\text{eff}}-1}\,.
\end{equation}
This effective step width is related to the degree of correlation between nearby frequencies and can be understood as a rough estimate of the separation in frequency $\omega$ at which models become effectively independent. For the type of models under investigation $\Delta\omega_{\text{eff}}$ is independent of frequency and depends mainly on $l_{\text{max}}$ of the experiment.

\begin{figure}
 \centering
\includegraphics[width=0.8\columnwidth]{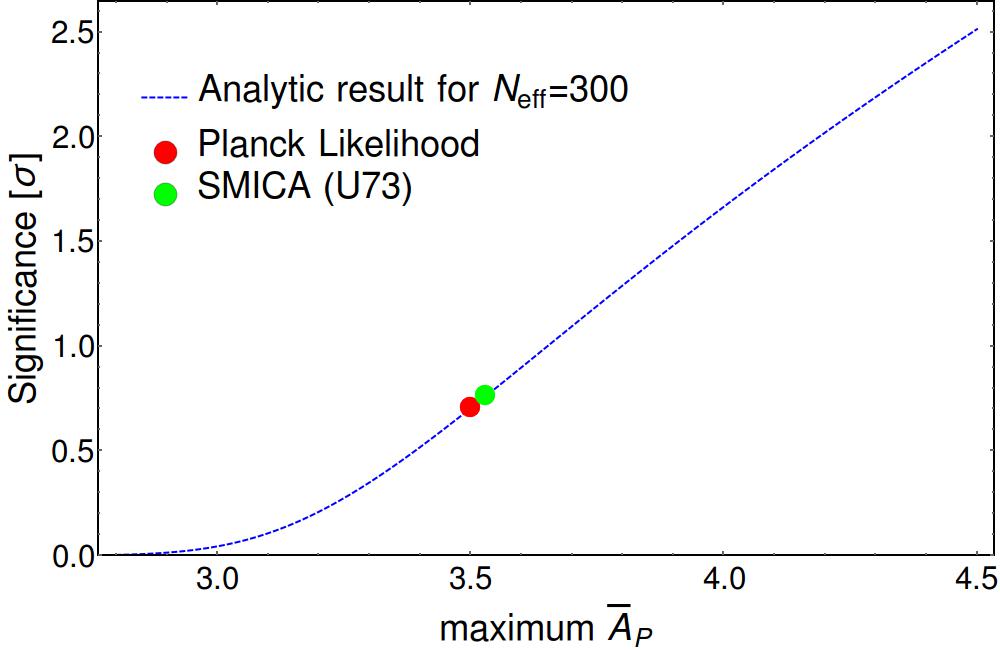}
\caption{Analytic result for the distribution of maximum significance feature model amplitudes in a survey with $N_{\text{eff}}=300$. The values obtained from the Planck Likelihood and the SMICA analysis are highlighted.}
\label{fig:sigind}
\end{figure}

In App.~\ref{app:statistics} we show that we have $\Delta\omega_{\text{eff}}\approx 13$ for a Planck-like set-up and, therefore, we arrive at $N_{\text{eff}}\approx300$ for our survey range. Note that this also justifies our choice of frequency step width $\Delta\omega=10$. Due to the strong correlations between frequencies separated by less than $\Delta\omega_{\text{eff}}$ this step width should be sufficiently small to resolve all peaks in the likelihood improvement. Figure~\ref{fig:sigind}%
shows the analytic relation between the maximum raw significance observed and the corresponding look-elsewhere-adjusted significance for $N_{\text{eff}}=300$ with the respective maximum values from the Planck Likelihood and SMICA analysis highlighted. The look-elsewhere-adjusted significances in both cases are clearly below the one sigma level. Loosely speaking this implies that one expects roughly every other random realisation of a featureless CMB to give rise to a maximum significance at least as big as what we observe in our CMB. Hence, we cannot conclude that the maximum likelihood improvements in the power spectrum on their own present convincing evidence for feature models at the respective frequencies.

\begin{figure}
 \centering
\includegraphics[width=0.8\columnwidth]{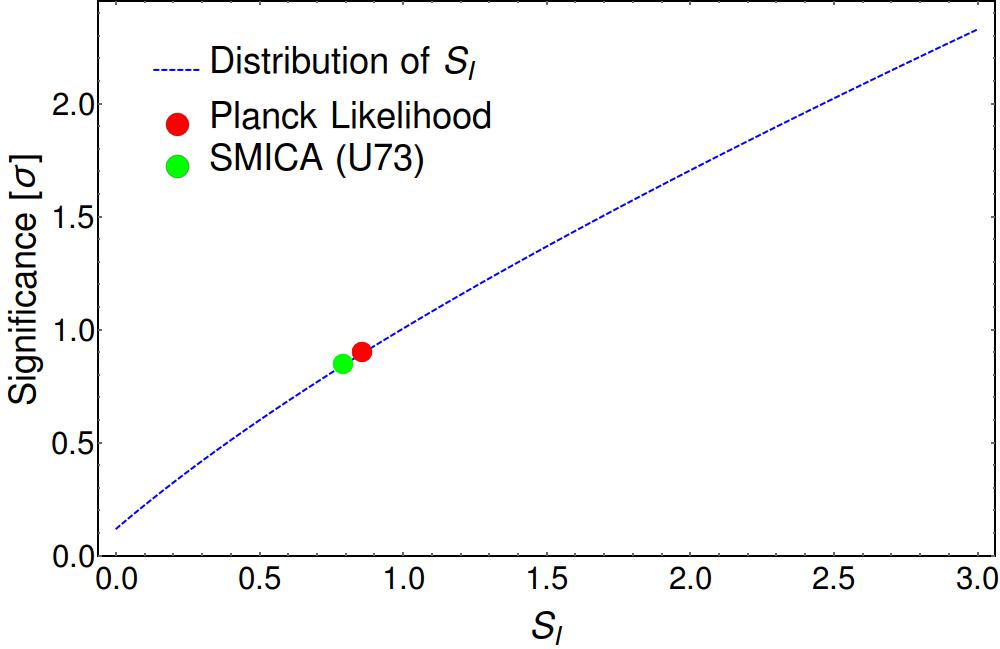}
\caption{Prediction of the distribution of the integrated statistic $S_I$. The values obtained from the Planck Likelihood and the SMICA analysis assuming $N_{\text{eff}}=300$ are highlighted.}
\label{fig:sigindint}
\end{figure}

\begin{figure}
 \centering
\includegraphics[width=\columnwidth]{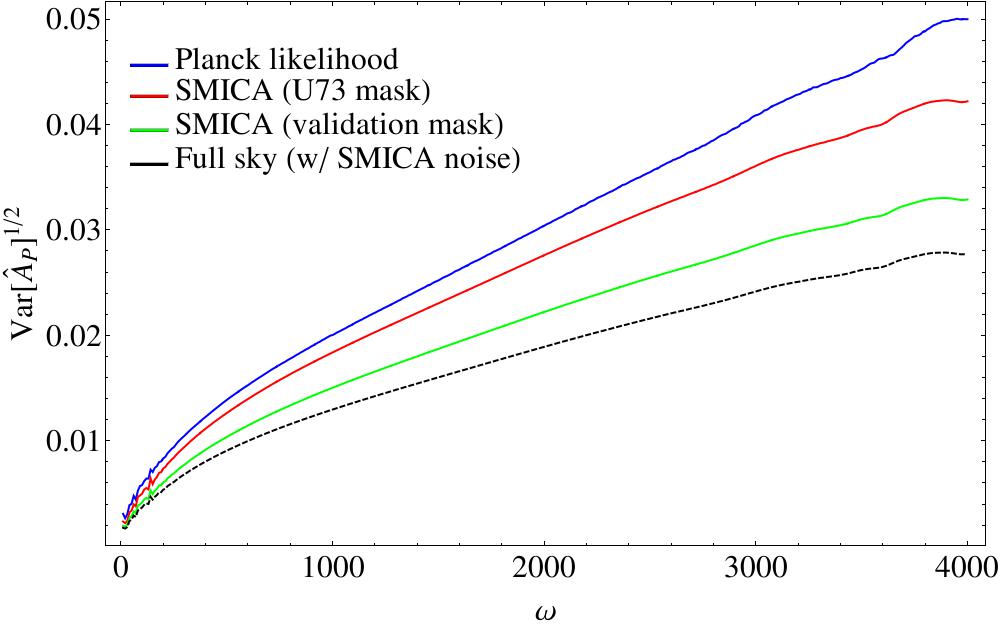}
\caption{The standard deviations of the amplitude measurements for the various analysis methods presented in this work. The Planck Likelihood uses the CL31 ($f_{\text{sky}}=0.31$) and CL49 ($f_{\text{sky}}=0.49$) masks while the SMICA analysis was carried out with an apodised version of the U73 mask ($f_{\text{sky}}=0.61$) and the validation mask ($f_{\text{sky}}=0.81$). The standard deviation evidently decreases with growing sky fraction or equivalently more included data.}
\label{fig:variances}
\end{figure}

In Ref.~\cite{Fergusson:psbsfeatures} a further test was suggested that addresses the possibility that the data could present evidence for feature models that give rise to modulations with multiple well-separated frequencies. This is different from simply looking at the maximum significance found in a survey as the height of all other peaks are taken into account as well. It was found that the integrated statistic $S_I$ given by
\begin{equation}\label{eq:sigindint}
S_I^2=\frac{\Delta\omega}{\Delta\omega_{\text{eff}}}\sum\limits_{\omega}2\,\text{Erf}^{-1}\left[\left(F_{\chi,\,2}\left(\bar{A}_{P,\omega}\right)\right)^{N_{\text{eff}}}\right]^2
\end{equation}
produces significances that agree well with a rigorous look-elsewhere analysis for multi-frequency models. Here, $\Delta\omega$ is the step width in frequency of the survey and the sum reduces to an integral over frequency in the $\Delta\omega\rightarrow 0$ limit. Figure~\ref{fig:sigindint}%
shows the distribution of $S_I$ taken from%
\footnote{We note that the fit stated in Ref.~\cite{Fergusson:psbsfeatures} contains a minor error. The CDF of the statistic $S_I$, $F_{S_I}(x)$, is well described by
\begin{equation}
F_{S_I}(x)=1-\exp\left(-(a\,x^2+b\,x+c)\right)
\end{equation}
where $c=0.102$ as in Ref.~\cite{Fergusson:psbsfeatures} but $a=0.108$ (rather than $a=0.092$) and $b=0.950$ (rather than $b=0.876$). We use the correct values in this work even though the small changes make little difference in the cases considered here.} %
Ref.~\cite{Fergusson:psbsfeatures} with the values obtained from the Planck Likelihood and the SMICA analysis highlighted. Again, both results are below the one sigma level indicating that the survey results are consistent with a random realisation of a featureless Gaussian CMB.

To sum up, the results above indicate that neither the maximum likelihood improvements nor the abundance of further large peaks present convincing evidence for a detection of features in the power spectrum based on the Planck Likelihood or the SMICA map. Of course, this does not exclude the possibility that the primordial power spectrum exhibits oscillatory features and with the inclusion of more data a different outcome with a positive detection might be reached.

There are several ways to make progress at this stage. In Sec.~\ref{subsec:Results-combined} we will combine our survey with bispectrum results. As feature models should also produce signatures in the bispectrum this can provide further evidence. Without invoking the bispectrum one can only include more data, for example the Planck polarisation data, to lower the error bars on feature model amplitudes%
\footnote{See for example the discussions in Refs.~\cite{Mortonson:PolFeatures,Miranda:PolarizationPredictions,Fergusson:psbsfeatures} and references therein.}. %
Without the polarisation data at hand, the only way to increase the amount of data is the inclusion of a larger sky fraction. This is not possible in the framework of the Planck Likelihood as, by construction, it only operates on the cleanest parts of the sky where foregrounds can be modelled as effective contributions to the power spectrum.

\begin{figure*}
 \centering
\includegraphics[width=\textwidth]{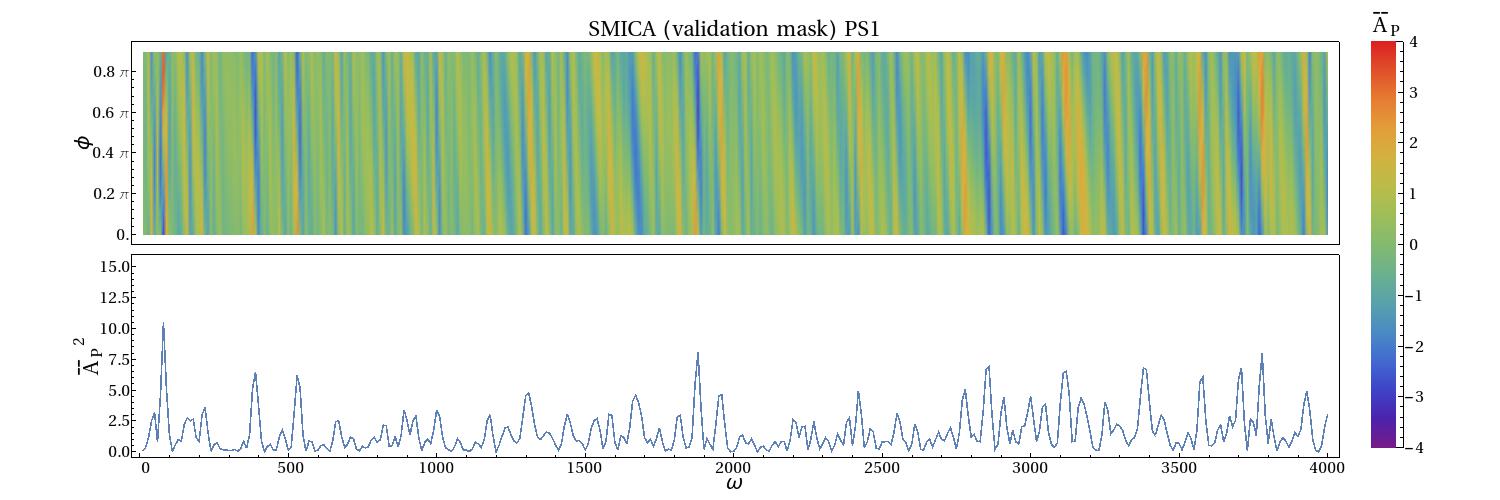}
\caption{Amplitudes $\bar{A}_P$ for the template PS1 up to $\omega=4000$ based on the SMICA map masked with an apodised version of the SMICA validation mask with $f_{\text{sky}}=0.81$. The bottom panel shows the maximum $\bar{A}_P^2$ at a given $\omega$, corresponding to the maximum likelihood improvement at that frequency.}
\label{fig:Smicavalo4000}
\end{figure*}

However, the SMICA component separation algorithm produces valid%
\footnote{As measured by a somewhat subjective criterion as described in Ref.~\cite{Ade:2013CompSep} and references therein.} %
results on more than 80\% of the sky, defined by the validation mask%
\footnote{The U73 mask is the union of the validation masks of the four component separation algorithms used in the Planck analysis. Some of these have significantly smaller sky fractions resulting in the smaller sky fraction of the union mask.} % 
of the algorithm. The validation mask was published as part of the 2013 Planck data release and we constructed an apodised version of it with $f_{\text{sky}}=0.81$. The larger sky fraction evidently leads to a significant reduction in error bars as can be seen in Fig.~\ref{fig:variances}.
If the large results we observe at various frequencies are in fact signatures of real features of the primordial power spectrum, we expect their significances to rise. This expectation obviously necessitates that we trust the SMICA algorithm over the region previously excluded by the U73 mask, but not by the validation mask. The results obtained using the validation mask in an otherwise identical analysis that led to the SMICA results in Fig.~\ref{fig:ampso4000} are presented in Fig.~\ref{fig:Smicavalo4000}.%
The large peaks observed in Fig.~\ref{fig:ampso4000} shrink rather than grow. This leads us to conclude that either the regions that were previously excluded are not faithful representations of the actual CMB due to a failure of the component separation algorithm to isolate the CMB component or the large signals are simply the result of fitting the scatter of the $\hat{C}_l$ by chance.

\subsubsection{Including the sharp feature scaling: results for the template PS2}
\label{subsubsec:PS2survey}
As pointed out in Sec.~\ref{sec:featuremodels}, the template PS1 given by Eq.~\eqref{eq:barePS} that we used in the previous section is not entirely appropriate when looking for the signatures of sharp features in the power spectrum. For these kinds of features the sine component of the signal is generically suppressed by a factor of $1/(\omega k)$ giving rise to the template PS2, Eq.~\eqref{eq:modPS}. Hence, it is interesting to see whether the large peaks observed in the previous section carry over to this case.

Before presenting the results we would like to draw attention to an important qualifier. The way we set up the search for signatures of feature models places some restrictions on the type of sharp features we can look for. First of all, the templates are only supposed to capture the behaviour accurately for $\omega k\gg 1$. As explained in Sec.~\ref{sec:featuremodels} for the template PS1 and frequencies $\omega\gg 100$ this is never really a problem as all the S/N comes from wavenumbers $k$ satisfying this condition. While this conclusion carries over to the cosine part of the PS2 template, the decaying sine part is more problematic. In particular, those models of the PS2 template that have mostly a decaying sine component ($\phi\sim0$ or $\phi\sim\pi$) might in fact receive a non-negligible contribution to their S/N that comes from wavenumbers $k$ that violate $\omega k\gg 1$. The SMICA analysis above discards multipoles with $l<50$ so that for $\omega\gg 100$ this region is not included in the estimates. To achieve the same in the Planck analysis we discard the low-$l$ likelihood and only use the Planck high-$l$ likelihood for the PS2 survey. Then, just like in the SMICA analysis, only multipoles with $l\ge 50$ are taken into account. For those models of the PS2 template that receive significant S/N contributions from $l<50$ (i.e.\ those with $\phi\sim0$ or $\phi\sim\pi$) this is evidently a suboptimal solution and an analysis that uses templates that capture the exact low-$l$ behaviour would produce better constraints on these models.

A further closely related limitation comes from the fact that we assume that the primordial power spectrum is linear in the feature model amplitude. While this should be a good approximation for amplitudes up to $\mathcal{O}(0.1)$, it clearly breaks down for amplitudes that would imply order unity modulations. In extreme cases these would lead to negative values of the primordial power spectrum. The correct templates in these cases have to be obtained from an appropriate non-linear GSR approximation \cite{Dvorkin:GSR}. Again, this problem never occurs when looking for the template PS1 as can be easily seen from the standard deviations on the amplitude $\hat{A}_P$ shown in Fig.~\ref{fig:variances}. For these models the modulations of the primordial power spectrum never become of order unity%
\footnote{Unless we detect a signal at a significance of $\gtrsim 20$ sigma. If such a strong signature was detected, it should be studied separately using exact predictions for the modulations induced by corresponding feature models. Unfortunately, we never observe a signal that is even remotely close to this level of significance.}. %
However, for the decaying sine component of the template PS2 the sharp rise of the modulation amplitude towards small $k$ can be problematic. The only models that are significantly affected are those with $\phi\sim0$ or $\phi\sim\pi$ that contain a large decaying sine contribution. For large frequencies $\omega\gg 2000$ a highly significant result exceeding four sigma would imply order unity modulations at small $k$ for these models. Such results should be interpreted with care as non-linear corrections to the template might have non-negligible effects on the results.

Summing up, when studying the figures in this section care should be taken when interpreting the results near $\phi=0$ or $\phi=\pi$ as the results might not accurately reflect the signature of a corresponding sharp feature or at least can be significantly improved by an analysis that takes into account the correct small-$k$ behaviour.

\begin{figure*}
 \centering
\includegraphics[width=\textwidth]{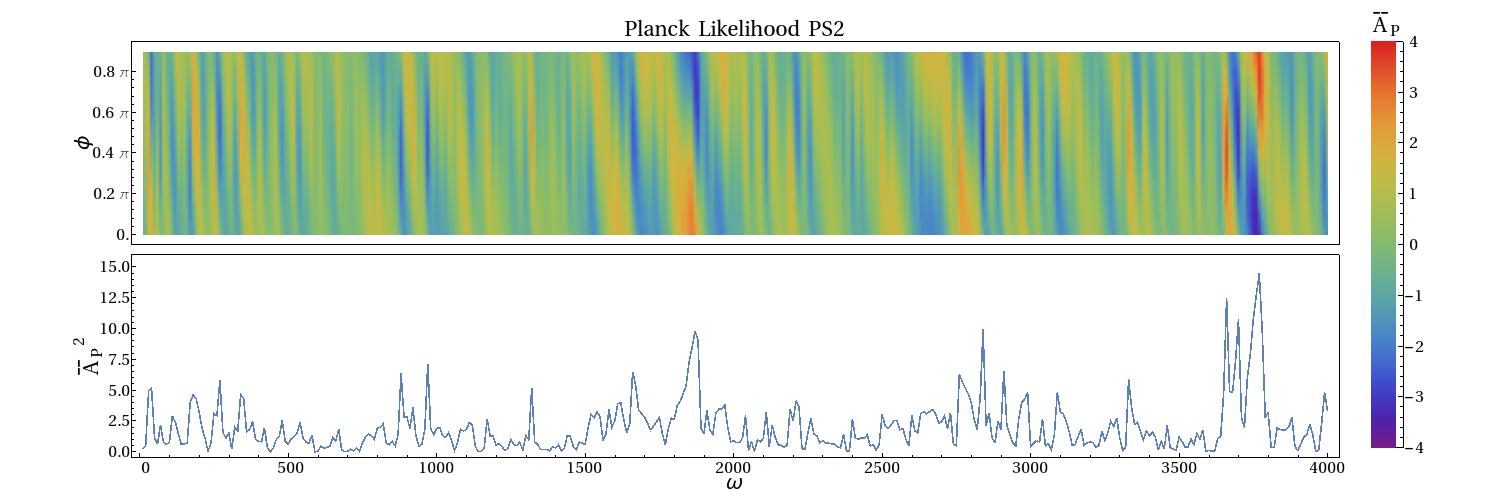}
\includegraphics[width=\textwidth]{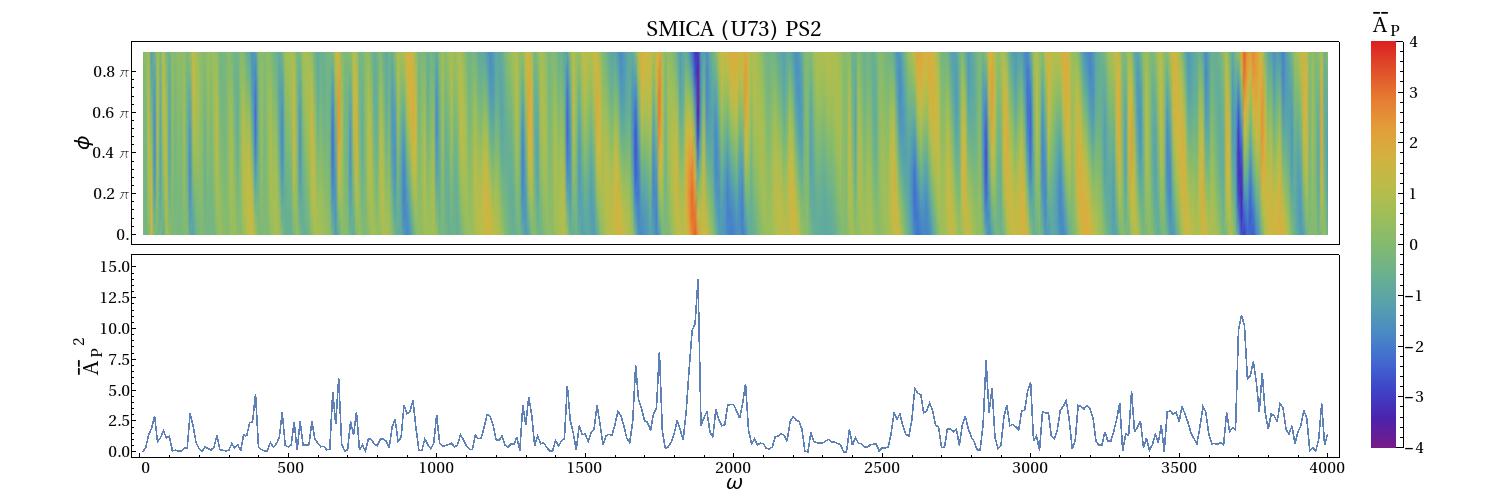}
\caption{Amplitudes $\bar{A}_P$ for the template PS2 (decaying sine) up to $\omega=4000$ based on the Planck Likelihood (top) and the SMICA map masked with an apodised version of the U73 mask with $f_{\text{sky}}=0.61$ (bottom). In each plot the bottom panel shows the maximum $\bar{A}_P^2$ at a given $\omega$, corresponding to the maximum likelihood improvement at that frequency.}
\label{fig:ampsmodo4000}
\end{figure*}

Figure~\ref{fig:ampsmodo4000}%       
shows the results for the template PS2 for both the Planck Likelihood and the SMICA map masked with an apodised version of the U73 mask. We see that the large signal in the SMICA analysis at $\omega\sim1880$ slightly benefits from the inclusion of the decaying sine and is now observed at \num{3.7} sigma, whereas the significance remains basically unchanged in the Planck Likelihood. The best fit is still at $\phi\sim0.7\pi$ as in Fig.~\ref{fig:ampso4000}, so is dominated by the unsuppressed cosine and should be well in the parameter region where our analysis is valid.

On the other hand, the peak at $\omega\sim3710$ shifts slightly down in significance and is seen at \num{3.3} sigma in both searches. This could imply that an interpretation in terms of a sharp feature signal is potentially problematic. The largest peak ($\omega\sim3770$) in the Planck Likelihood analysis is now detected at \num{3.8} sigma corresponding to a likelihood improvement of $2\Delta\log{\mathcal{L}}\approx 14$.  However, we emphasise that it receives a considerable contribution from the decaying sine as it is seen at $\phi\sim 0.8\pi$, so an interpretation in terms of a sharp feature might require further investigation. Finally, the significance of the peak at $\omega\sim2840$ in the Planck Likelihood remains unchanged at \num{3.2} sigma.

\begin{figure*}
 \centering
\includegraphics[width=.8\textwidth]{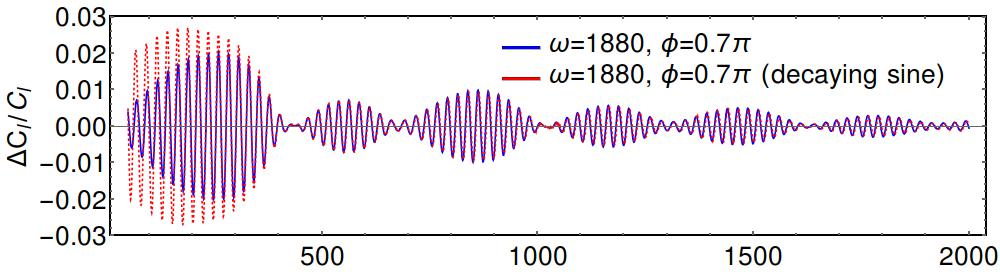}
\includegraphics[width=.8\textwidth]{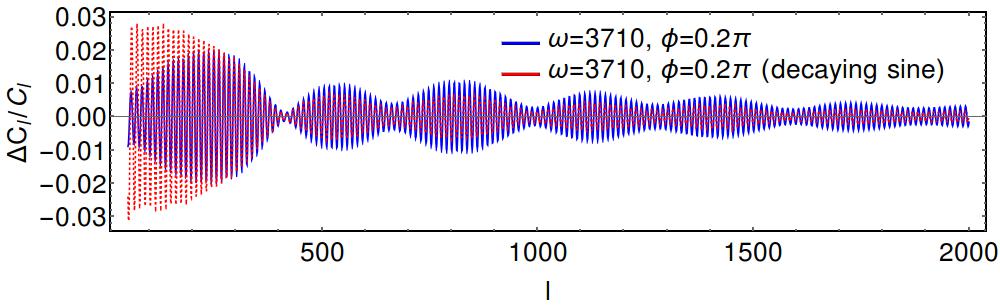}
\caption{Best-fit models based on the SMICA analysis for the template PS1 at $\omega=1880$ with $\phi=0.7\pi$ and $\omega=3710$ with $\phi=0.2\pi$, respectively, with the corresponding fits of the template PS2.}
\label{fig:deltaClBF}
\end{figure*}

Figure~\ref{fig:deltaClBF}%
displays the best-fit models for the SMICA analysis at $\omega\sim1880$ and $\omega\sim3710$, respectively. Also plotted are the corresponding PS1 templates. One can clearly see how the decaying sine alters the low-$l$ behaviour. At high $l$ the best-fit PS1 and PS2 templates are identical up to a shift in phase and amplitude of the modulations.

The statistics to discuss the look-elsewhere-adjusted significance were developed focusing on the template PS1 \cite{Fergusson:psbsfeatures}. There are differences in the correlation structure of the templates PS1 and PS2 as can be seen directly by comparison of Figs.~\ref{fig:ampso4000} and \ref{fig:ampsmodo4000}. While the cross section through the density plot at $\phi=0.5\pi$ must be the same (the templates are identical in this case), the decaying sine is more correlated in frequency which results in broader peak patterns around $\phi=0$ and $\phi=\pi$ %
\footnote{Heuristically, the suppression of oscillations with increasing $l$ can be thought of as effectively introducing a lower $l_{\text{max}}$ beyond which oscillations are negligible. This implies a larger effective step width $\Delta\omega_{\text{eff}}$ and, hence, broader peak patterns (cf.\ App.~\ref{app:statistics}).}. %
With this caveat in mind it should be clear from Fig.~\ref{fig:sigind} that none of the results will give rise to significant improvements once the look-elsewhere effect is taken into account. In particular, the highest result with $\bar{A}_P\approx 3.8$ at $\omega\sim3770$ in the Planck Likelihood analysis is roughly at the one sigma level.

\begin{figure*}
 \centering
\includegraphics[width=\textwidth]{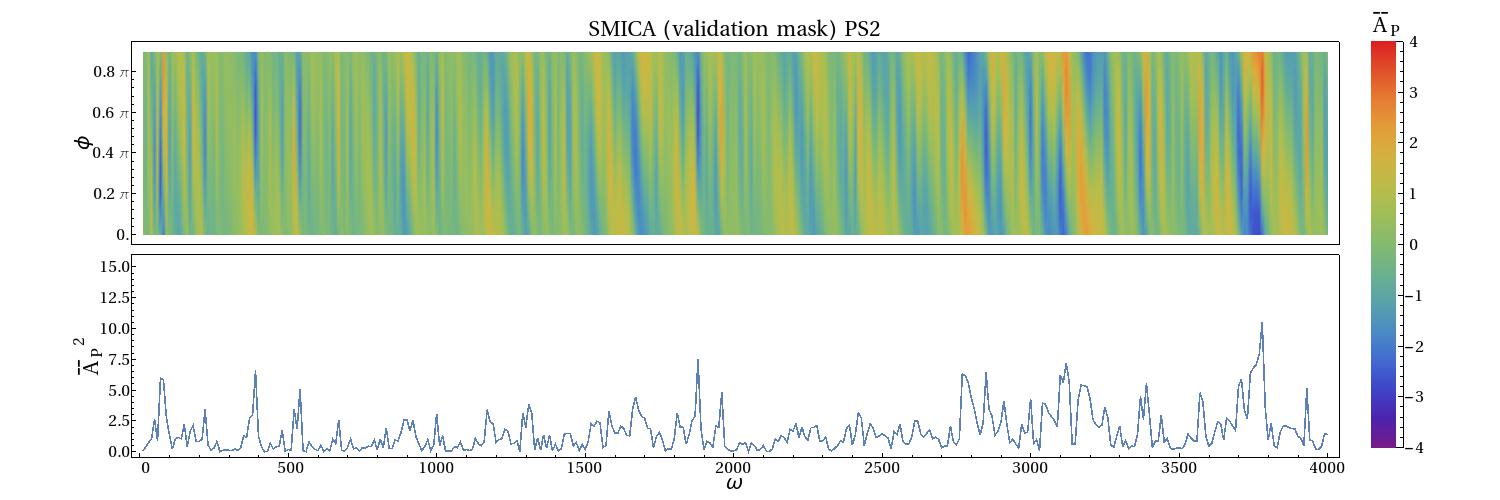}
\caption{Amplitudes $\bar{A}_P$ for the template PS2 (decaying sine) up to $\omega=4000$ based on the SMICA map masked with an apodised version of the validation mask with $f_{\text{sky}}=0.81$. The bottom panel shows the maximum $\bar{A}_P^2$ at a given $\omega$, corresponding to the maximum likelihood improvement at that frequency.}
\label{fig:ampsmodvalo4000}
\end{figure*}

Finally, repeating the same analysis for the validation mask as in the previous section does not lead to an expected increase in significance as can be seen in Fig.~\ref{fig:ampsmodvalo4000}.
Both large results observed in the SMICA analysis above (cf.\ Fig.~\ref{fig:ampsmodo4000}) decrease in significance. In particular, the strongest result seen in Fig.~\ref{fig:ampsmodo4000} at $\omega\sim1880$ drops below the three sigma level. Again, we emphasise that the meaningfulness of these results depends on the reliability of the component separation method over the entire region of the sky not excluded by the validation mask.

\subsection{WMAP bispectrum survey}
\label{subsec:Results-Bispectrum}
\subsubsection{Bare sine and cosine: results for the template BS1}
\begin{figure*}
 \centering
\includegraphics[width=\columnwidth]{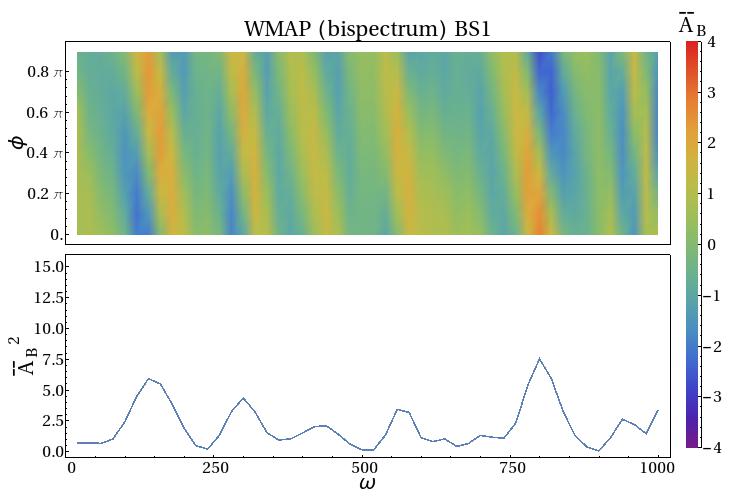}
\includegraphics[width=\columnwidth]{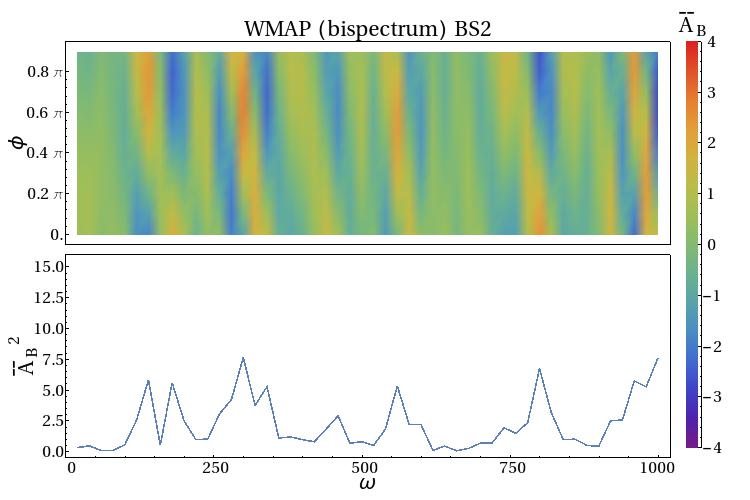}
\caption{Amplitudes $\bar{A}_B$ for the template BS1 (left) and BS2 (right) up to $\omega=1000$ as obtained from the WMAP data. In each plot the bottom panel shows the maximum $\bar{A}_B^2$ at a given $\omega$.}
\label{fig:ampsBSWMAPo1000}
\end{figure*}

The results for the template BS1 extracted from WMAP 9-yr data up to $\omega=1000$ are shown on the left of Fig.~\ref{fig:ampsBSWMAPo1000}. 
As in the corresponding plot for the power spectrum, the upper panel shows a density plot of the normalised amplitudes $\bar{A}_B$, while the lower panel shows the maximum measured $\bar{A}_B^2$ at a given frequency. The highest peak is found at $\omega\sim800$ with a significance of about \num{2.8} sigma.

\begin{figure}
 \centering
\includegraphics[width=.8\columnwidth]{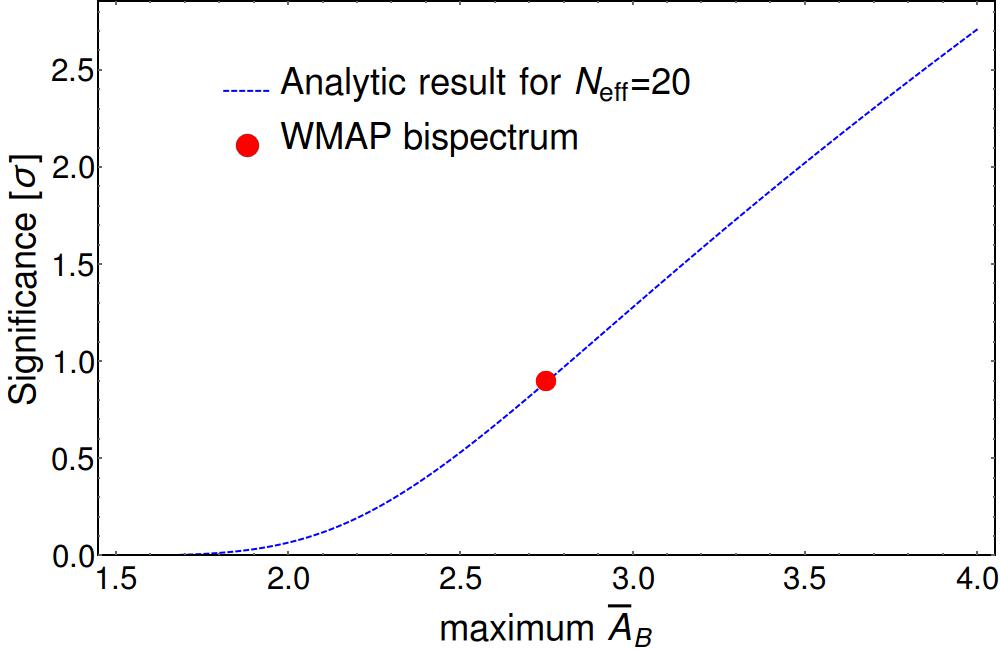}
\caption{Analytic model for the distribution of the maximum significance feature model amplitude $\hat{A}_B$ in a survey with $N_{\text{eff}}=20$. The value obtained from the WMAP data is highlighted.}
\label{fig:sigindBS}
\end{figure}

To judge whether the observed peaks are at a significant level after the look-elsewhere effect has been taken into account, we can again make use of the statistic Eq.~\eqref{eq:sigind} with an appropriate choice of $N_{\text{eff}}$. For an analysis with $l_{\text{max}}=600$ and WMAP noise level it is shown in App.~\ref{app:statistics} that we have%
\footnote{Note that the ratio of the values of $\Delta\omega_{\text{eff}}$ for WMAP and Planck, 50 and 13 respectively, is in reasonable agreement with a rough estimate of an effective $l_{\text{max}}$ of the two experiments, $\sim 600$ and $\sim 2000$, as one would expect based on how oscillatory templates should be correlated on a given domain (cf.\ App.~\ref{app:statistics}).} %
$\Delta\omega_{\text{eff}}\approx50$. This implies that for a survey covering a frequency range up to $\omega=1000$ we have $N_{\text{eff}}\approx20$. The significance of a given $\bar{A}_B$ based on Eq.~\eqref{eq:sigind} is presented in Fig.~\ref{fig:sigindBS}%
and the value determined from the WMAP survey presented in this work is highlighted. The measured amplitudes are evidently below the one sigma level implying that none of the observed peaks on their own present evidence for an oscillation in the bispectrum.

\begin{figure}
 \centering
\includegraphics[width=.8\columnwidth]{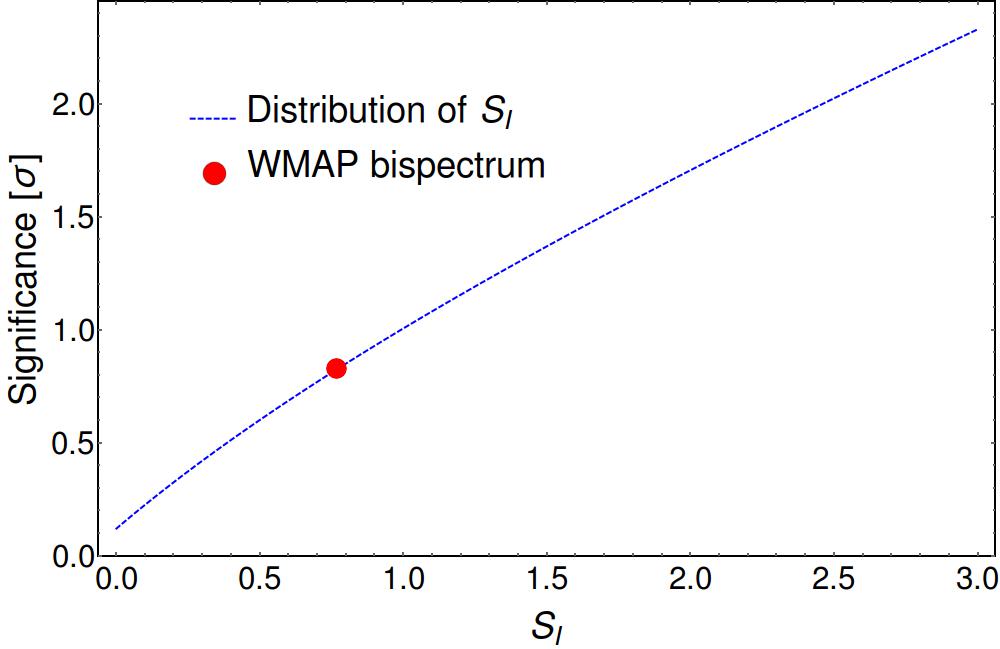}
\caption{Distribution of the integrated statistic $S_I$. The value obtained from the WMAP bispectrum survey assuming $N_{\text{eff}}=20$ is highlighted.}
\label{fig:sigindintBS}
\end{figure}

As in the power spectrum analysis above, we also consider the integrated statistic $S_I$ from Eq.~\eqref{eq:sigindint} evaluated on the WMAP bispectrum data which is shown in Fig.~\ref{fig:sigindintBS}.
Just as in the case of the single peak statistic, we arrive at a value below the one sigma level meaning that the abundance of large peaks is entirely consistent with a Gaussian featureless CMB.

\subsubsection{Including the sharp feature scaling: results for the template BS2}
As in the case of the power spectrum, we pointed out in Sec.~\ref{sec:featuremodels}, that the bispectrum template BS1 given by Eq.~\eqref{eq:bareBS} that we used in the previous section does not correctly capture the sharp feature scaling of the shape. The sine component of the signal generically comes with a factor of $\omega K$ while the cosine component is multiplied by a factor of $(\omega K)^2$. We included these scalings in our template BS2, Eq.~\eqref{eq:modBS}. We go on to investigate whether the inclusion of these scalings changes the results from the previous section.

The results for the template BS2 extracted from WMAP 9-yr data up to $\omega=1000$ are shown on the right of Fig.~\ref{fig:ampsBSWMAPo1000}. There are clear similarities to the results for the template BS1 without the correct scaling on the left. This is particularly true for the sine component ($\phi\sim 0$ and $\phi\sim\pi$) showing very correlated patterns in the two plots. However, the cosine component seems to be strongly affected by the inclusion of the $(\omega K)^2$ scaling causing qualitative differences. Generally, the correlation width in frequency seems to have decreased.

None of the peaks have substantially gained in significance due to the inclusion of the sharp feature scaling with the maximum amplitude still being $\bar{A}_B\approx 2.8$. If we use the same statistic as for the BS1 template, it should be clear from Fig.~\ref{fig:sigindBS} that the results do not exceed the one sigma level. Again with the caveat in mind that the distribution of the maximum amplitudes might be slightly affected due to the different correlation structure of the template BS2 (cf.\ Sec.~\ref{subsubsec:PS2survey}), we are led to conclude that no convincing evidence for the presence of sharp features in the WMAP bispectrum alone can be extracted in this frequency range.

\subsection{Combined power spectrum and bispectrum survey}
\label{subsec:Results-combined}
In Ref.~\cite{Fergusson:psbsfeatures} a natural statistic was introduced to identify evidence for feature models in a combined survey. We expect feature models to exhibit the same frequency $\omega$ in both the power spectrum and the bispectrum (cf.\ App.~\ref{app:sharpfeatures}). However, other parameters such as the relative amplitude of the signal in the bispectrum compared to the power spectrum, $A_B/A_P$, are very model-dependent. Thus, it is sensible to construct the maximum significance joint amplitude estimate $\bar{A}$ at a given $\omega$ by maximising over the ratio $A_B/A_P$ (and the phases of the oscillation, $\phi_P$ and $\phi_B$). This results in \cite{Fergusson:psbsfeatures}
\begin{equation}
\bar{A}=\left(\bar{A}_P^2+\bar{A}_B^2\right)^{\frac{1}{2}}\,,
\end{equation}
where $\bar{A}_P$ and $\bar{A}_P$ are the individual (normalised) power spectrum and bispectrum amplitudes at that frequency.

An analytic model for the distribution of the maximum joint amplitude estimate found in a combined survey, analogous to Eq.~\eqref{eq:sigind} that holds in the case of an individual survey, was presented in Ref.~\cite{Fergusson:psbsfeatures}. The essential difference is that in this case the CDF of the $\chi$-distribution with two degrees of freedom is replaced by the corresponding distribution with four degrees of freedom giving
\begin{equation}\label{eq:sigcomb}
S=2^{\frac{1}{2}}\text{Erf}^{-1}\left[\left(F_{\chi,4}\left(\bar{A}\right)\right)^{N_{\text{eff}}}\right]\,.
\end{equation}
Similarly, for a combined survey the integrated statistic $S_I$, analogous to Eq.~\eqref{eq:sigindint}, is given by
\begin{equation}\label{eq:sigcombint}
S_I^2=\frac{\Delta\omega}{\Delta\omega_{\text{eff}}}\sum\limits_{\omega}2\,\text{Erf}^{-1}\left[\left(F_{\chi,4}\left(\bar{A}_{\omega}\right)\right)^{N_{\text{eff}}}\right]^2\,.
\end{equation}
In both of these definitions an appropriate choice of $N_{\text{eff}}$ is required. In the case of combining surveys with identical $\Delta\omega_{\text{eff}}$ as in Sec.~\ref{subsubsec:WMAPplusWMAP} the choice is obvious. The choice of $N_{\text{eff}}$ for a combined survey with different $\Delta\omega_{\text{eff}}$ is discussed in Sec.~\ref{subsubsec:WMAPplusPlanck}.

\subsubsection{WMAP bispectrum and WMAP power spectrum}
\label{subsubsec:WMAPplusWMAP}
\begin{figure}
 \centering
\includegraphics[width=\columnwidth]{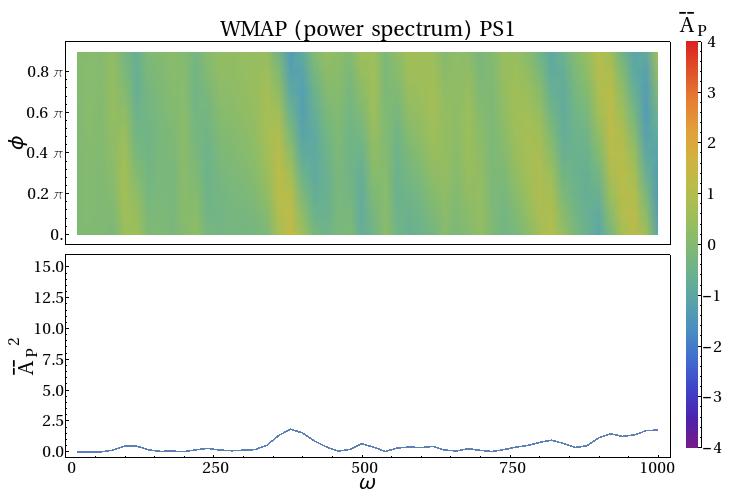}
\caption{Amplitudes $\bar{A}_P$ for the template PS1 up to $\omega=1000$ based on the WMAP likelihood. As in previous figures, the bottom panel shows the maximum $\bar{A}_P^2$ at a given $\omega$, corresponding to the maximum likelihood improvement at that frequency.}
\label{fig:ampsPSWMAPo1000}
\end{figure}

The results for the template PS1 up to $\omega=1000$ as extracted from the WMAP likelihood using the efficient methods described in Sec.~\ref{subsubsec:PlanckLike} are shown in Fig.~\ref{fig:ampsPSWMAPo1000}.
They agree very well with the corresponding figure in Ref.~\cite{Meerburg:2013SearchOscP1} as expected.

Before proceeding to perform a combined analysis we would like to point out that the WMAP power spectrum results below $\omega=1000$ are curiously low with a maximum of $\bar{A}_P\approx1.4$. We remind the reader that we expect $N_{\text{eff}}=20$ for WMAP and the given frequency range as discussed above. Simply employing the analytic model for the distribution of the maximum amplitude from Ref.~\cite{Fergusson:psbsfeatures} we find that such a low maximum should only occur roughly once in \num{10000} realisations which would make the absence of peaks in this region a four-sigma anomaly. The analytic distribution is not entirely accurate for judging very low-significance results as it was not designed for this purpose. Closer inspection using MC sampling shows that this anomaly is likely around the three sigma level. At the present stage it is unclear whether this is simply due to an unlikely realisation of the low-frequency scatter around $\Lambda$CDM or due to a step in the data processing that systematically eliminates low-frequency oscillations.

\begin{figure}
 \centering
\includegraphics[width=\columnwidth]{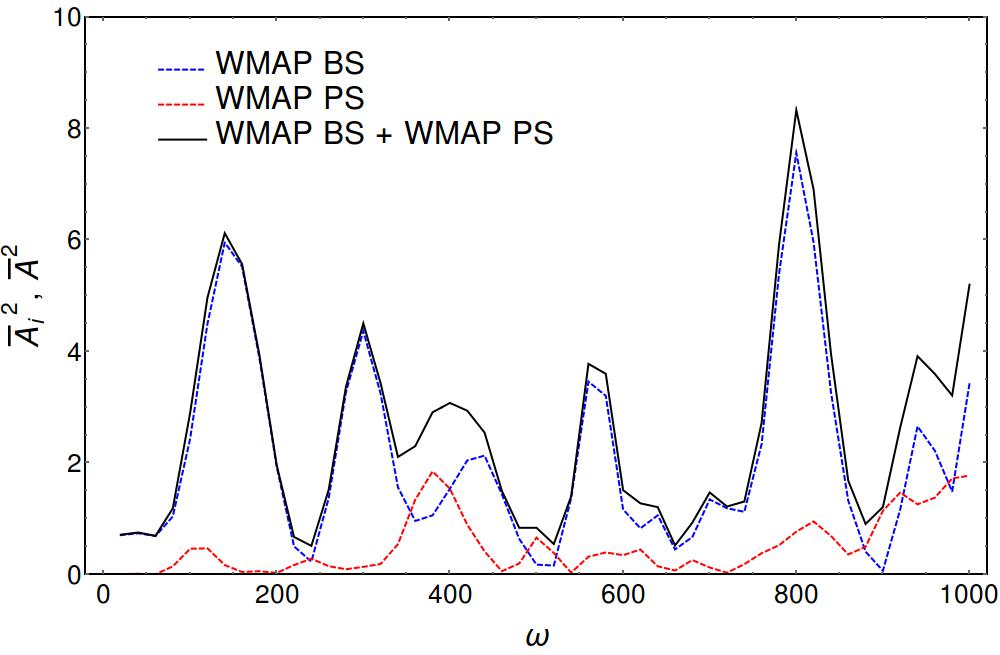}
\caption{Individual amplitudes $\bar{A}^2_P$ and $\bar{A}^2_B$ and combined amplitude estimates $\bar{A}^2$ at a given $\omega$ obtained from combining the WMAP bispectrum survey with the WMAP power spectrum survey up to $\omega=1000$.}
\label{fig:ampscombWMAPWMAP}
\end{figure}

The values of the squared combined amplitude estimates $\bar{A}^2$ obtained from combining the WMAP bispectrum survey with the WMAP power spectrum up to $\omega=1000$ are shown in Fig.~\ref{fig:ampscombWMAPWMAP}.
Due to the absence of large peaks in the WMAP power spectrum in this region most of the contributions to $\bar{A}^2$ come from the bispectrum. Furthermore, the small peaks that can be seen in the power spectrum do not match those in the bispectrum. Hence, we do not expect this combined survey to present us with more significant evidence.

\begin{figure}
 \centering
\includegraphics[width=0.8\columnwidth]{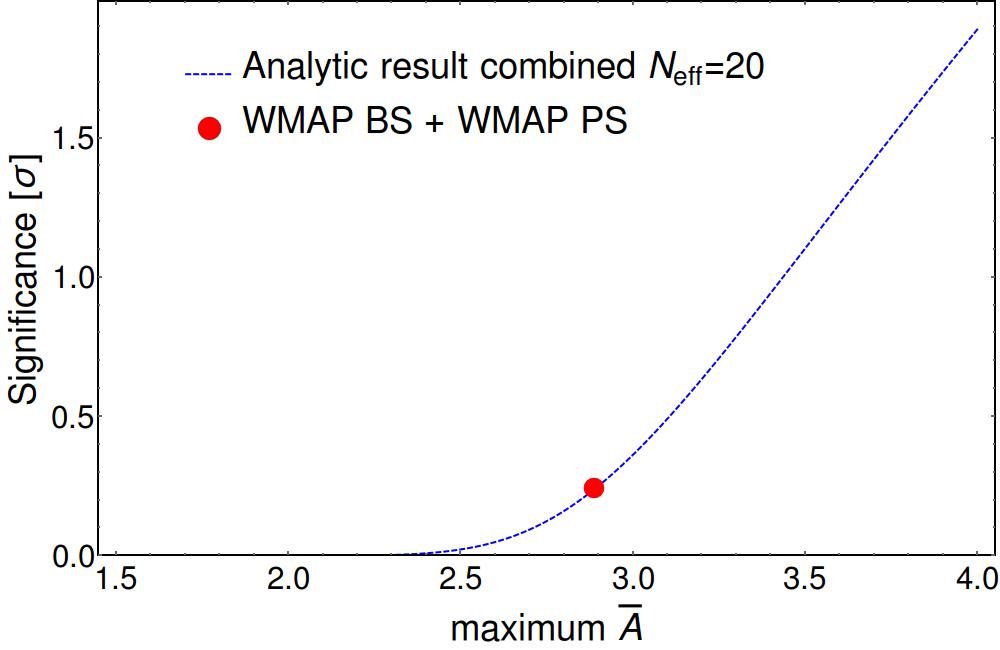}
\caption{Analytic model for the distribution of the maximum amplitude estimate $\bar{A}$ in a combined survey with $N_{\text{eff}}=20$. The value for the combined WMAP bispectrum and WMAP power spectrum analysis is highlighted.}
\label{fig:sigindcombWMAPWMAP}
\end{figure}

\begin{figure}
 \centering
\includegraphics[width=0.8\columnwidth]{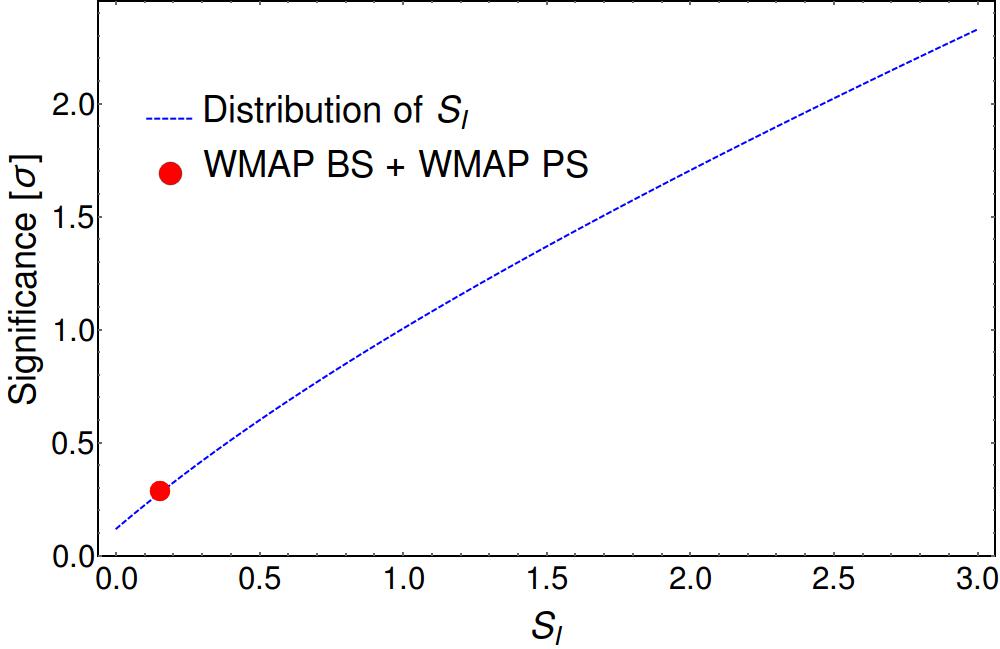}
\caption{Distribution of the integrated statistic $S_I$. The values obtained from the combined WMAP bispectrum and WMAP power spectrum survey assuming $N_{\text{eff}}=20$ is highlighted.}
\label{fig:sigindintcombWMAPWMAP}
\end{figure}

The statistics discussed above confirm this. The result from the statistic for the maximum joint amplitude estimate, Eq.~\eqref{eq:sigcomb}, is shown in Fig.~\ref{fig:sigindcombWMAPWMAP}.
It is well below the one sigma level and noticeably lower than the corresponding result for the bispectrum only analysis, Fig.~\ref{fig:sigindBS}. The result from evaluating the integrated statistic, Eq.~\eqref{eq:sigcombint}, is shown in Fig.~\ref{fig:sigindintcombWMAPWMAP}. 
Again, the result is low and dropped compared to its value for the bispectrum only analysis in Fig.~\ref{fig:sigindintBS}. The decrease in significance as judged by these statistics is due to the absence of large results in the power spectrum in this region and the mismatch of peaks in the power spectrum and bispectrum. The larger look-elsewhere effect that arises in a combined search with more parameters is in this case not matched by corresponding larger observed significances $\bar{A}$.

\subsubsection{WMAP bispectrum and Planck/SMICA power spectrum}
\label{subsubsec:WMAPplusPlanck}
\begin{figure*}
 \centering
\includegraphics[width=\columnwidth]{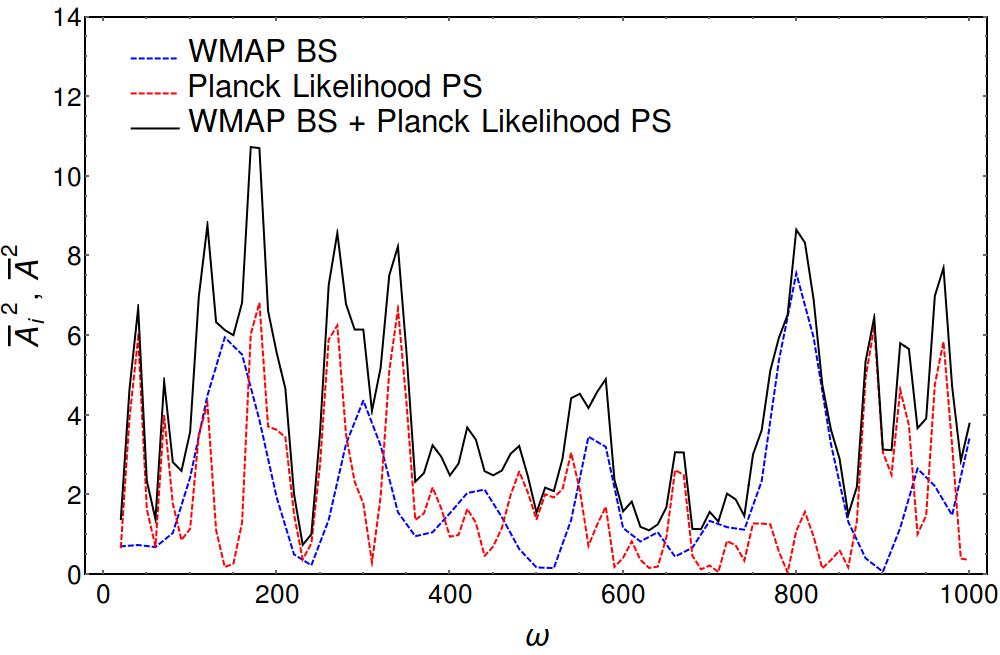}
\includegraphics[width=\columnwidth]{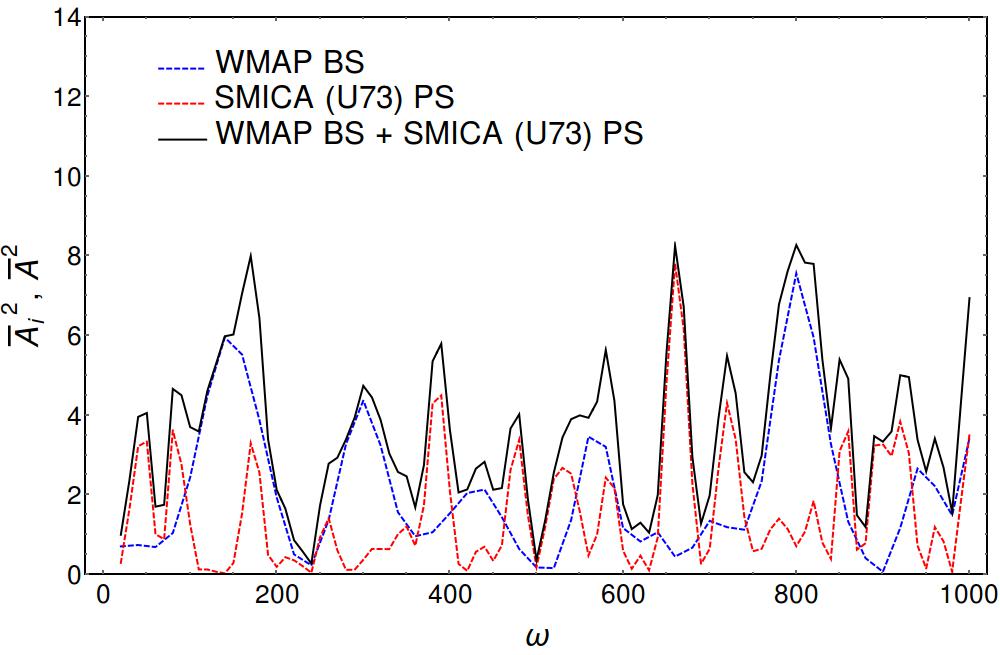}
\caption{Individual amplitudes $\bar{A}^2_P$ and $\bar{A}^2_B$ and combined amplitude estimates $\bar{A}^2$ at a given $\omega$ obtained from combining the WMAP bispectrum survey with either the Planck Likelihood (left) or the SMICA power spectrum survey (right) up to $\omega=1000$.}
\label{fig:ampscomb}
\end{figure*}

The values of $\bar{A}^2$ obtained from combining the WMAP bispectrum survey with the Planck Likelihood up to $\omega=1000$ are shown on the left of Fig.~\ref{fig:ampscomb}.
The corresponding results for the SMICA power spectrum survey are shown on the right. As $\bar{A}^2$ is simply the sum of the contributions from the individual surveys, it inherits the highly irregular shape with many local maxima.

Again, we would like to decide whether or not any of these joint amplitude estimates present significant evidence. In the present case, there is a further complication related to the fact that we are combining a WMAP bispectrum survey with Planck power spectrum surveys. These two types of surveys have different $\Delta\omega_{\text{eff}}$ due to their different noise levels and $l_{\text{max}}$. Hence, it is not immediately clear which value to plug into Eq.~\eqref{eq:domeffdef} to extract the correct value of $N_{\text{eff}}$. In App.~\ref{app:statistics} it is shown that Eq.~\eqref{eq:sigcomb} is an excellent model for the distribution of the maximum $\bar{A}$ if $N_{\text{eff}}$ is taken to be the arithmetic mean of the values of $N_{\text{eff}}$ for the individual surveys. Equivalently, the $\Delta\omega_{\text{eff}}$ of the combined survey is the harmonic mean of the values of $\Delta\omega_{\text{eff}}$ of the individual surveys. This gives $\Delta\omega_{\text{eff}}\approx 21$ for the combined survey. For a frequency range up to $\omega=1000$ this corresponds to $N_{\text{eff}}\approx 48$.

\begin{figure}
 \centering
\includegraphics[width=0.8\columnwidth]{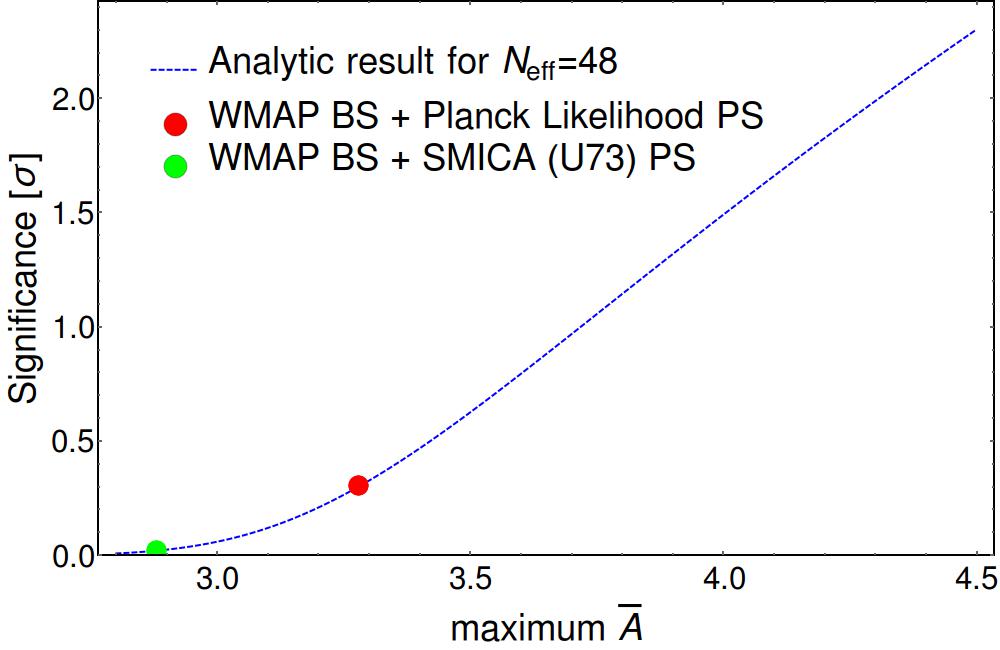}
\caption{Analytic model for the distribution of the maximum amplitude estimate $\bar{A}$ in a combined survey with $N_{\text{eff}}=48$. The values for the two combined surveys considered in this work are highlighted.}
\label{fig:sigindcomb}
\end{figure}

The corresponding distribution with the values for the two combined surveys highlighted is shown in Fig.~\ref{fig:sigindcomb}.
Both of these combined surveys give a result well below the one sigma level. Note that this result is a combination of the fact that the individual surveys show no significant peaks below $\omega=1000$ as is evident from Figs.~\ref{fig:ampso4000} and~\ref{fig:ampsBSWMAPo1000} and the fact that the largest peaks in this region occur at different frequencies and, hence, do not enhance each other. More precisely, there are peaks with $\bar{A}_P^2,\bar{A}_B^2\approx8$ in both the power spectrum and the bispectrum survey. If these were located at the same $\omega$, they would produce a joint estimate approaching the two sigma level.

\begin{figure}
 \centering
\includegraphics[width=0.8\columnwidth]{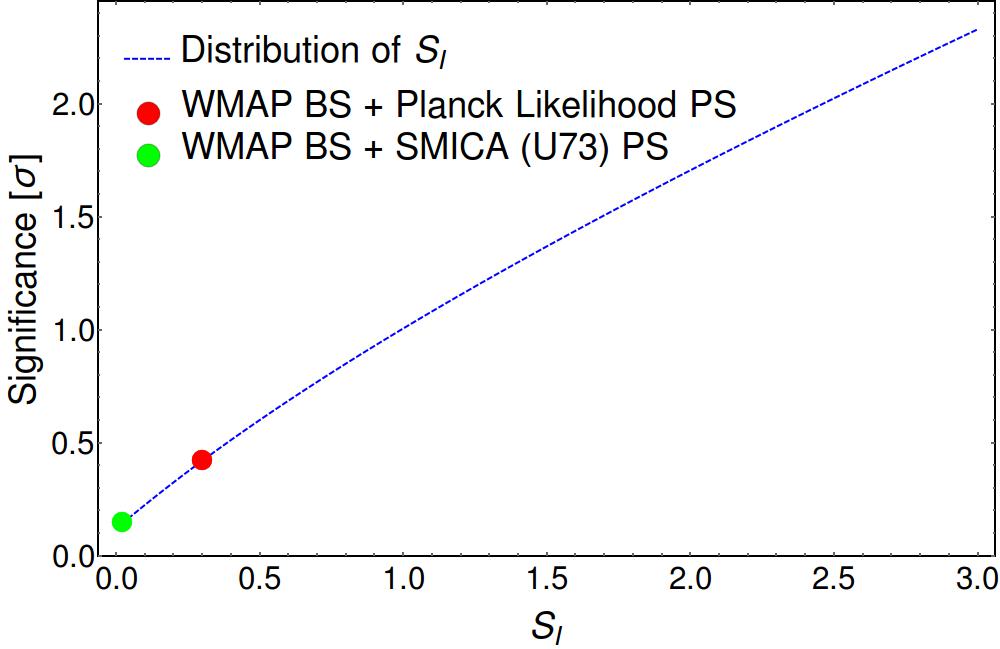}
\caption{Distribution of the integrated statistic $S_I$. The values obtained from the two combined surveys considered in this work assuming $N_{\text{eff}}=48$ are highlighted.}
\label{fig:sigindintcomb}
\end{figure}

Using the appropriate values for the combined survey, $\Delta\omega_{\text{eff}}\approx 21$ and $N_{\text{eff}}\approx 48$, we arrive at the results presented in Fig.~\ref{fig:sigindintcomb}.
The integrated statistic $S_I$ produces low values in both cases.

\section{Summary and Conclusions}
\label{sec:summconc}

In this paper, we undertook a thorough search for signatures of sharp features in Planck and WMAP9 data.  For the first time, we carried out searches in both the power spectrum and bispectrum simultaneously, employing well-defined look-elsewhere statistics to assess significances in a rigorous manner.

We developed highly efficient methods to scan the power spectrum for sharp oscillatory features with frequencies larger than the comoving sound horizon at LS, $\omega\gg140$. Demonstrating that in this case the only degeneracies of feature degrees of freedom with cosmological parameters are realised through uninteresting adjustments of the comoving distance to LS $\eta_{*}$, we argue that it is entirely sufficient to introduce a grid in frequency, keep the cosmological parameters fixed and only vary the feature model amplitude. In particular, varying $\eta_{*}$ does not produce bigger maximum likelihood improvements so that there is no risk of missing possibly interesting results adopting this simplified procedure. The only effect of a change is that frequencies close to a given peak in the likelihood improvement can benefit by an appropriate adjustment that changes the resulting effective oscillation in $C_l$ to match the one observed in the data.

Given that power spectrum likelihoods are very nearly Gaussian in the amplitude of feature models, the use of time consuming MCMC runs is not required and the best-fit amplitude and a corresponding significance can be extracted by fitting a Gaussian. We validated against a full MCMC analysis and found excellent agreement.

We employed these methods to scan the Planck Likelihood and also a likelihood based on the SMICA component separation maps for the signatures of sharp features. The latter has the advantage that, working on the assumption that the foreground cleaning is reliable, a larger sky fraction can be included providing in principle the strongest constraints on oscillatory features in the temperature power spectrum to date.

For our search we used both the phenomenological bare sine and cosine modulations (template PS1) and also included the correct sharp feature scaling (template PS2) up to $\omega=4000$. The Planck Likelihood scans in the case of the template PS1 agree very well with the corresponding results in Ref.~\cite{Meerburg:2013SearchOscP2} further validating our method. The SMICA map produces consistent results with comparably large likelihood improvements in the same places. To investigate further, we also used the SMICA validation mask for the analysis including a substantially larger sky fraction and, therefore, lowering the error bars by nearly a factor of two. Rather than gaining in significance the large peaks disappear. If we trust the cleaning procedure in regions of higher contamination, this should be interpreted as evidence that the large peaks are the result of fitting the scatter in the power spectrum estimates by chance and not a true signal.

The latter point of view is supported by the statistics developed in Ref.~\cite{Fergusson:psbsfeatures} to judge the look-elsewhere-adjusted significances of findings in feature model surveys. According to these statistics, neither the maximum significance nor the abundance of large peaks in the survey are in any way surprising with significances below the one sigma level. This implies that our realisation of the CMB is entirely consistent with a featureless primordial power spectrum. The inclusion of the sharp feature scaling has an effect on the results, but does not lead to very significant gains.

We went on to study the WMAP bispectrum up to frequencies $\omega=1000$. While various peaks can be identified, neither the phenomenological bare sine and cosine modulations nor the inclusion of the correct sharp feature scaling in the bispectrum give rise to significant results after look-elsewhere adjustment. To perform a combined search in the slightly simpler case of two surveys with the same effective frequency step width $\Delta\omega_{\text{eff}}$, we first used WMAP power spectrum data. In the process, we noticed that there is a curious absence of peaks in the power spectrum likelihood improvement over this frequency range constituting a roughly three sigma anomaly. With this in mind, it is not surprising that this joint analysis does not lead to any evidence for features.

Slightly generalising the statistics to allow for different $\Delta\omega_{\text{eff}}$, we combined the WMAP bispectrum data with Planck Likelihood and SMICA power spectrum surveys. Again, none of the results are at a statistically significant level. This is a combination of the fact that the individual surveys show no highly significant results, but also due to the fact that peaks do not occur at matching frequencies.

Summing up, neither the inclusion of more sky fraction in the framework of the SMICA analysis nor the inclusion of the correct sharp feature scaling produce power spectrum likelihood improvements that are significant after accounting for the look-elsewhere effect and can thus be interpreted as evidence for features. Invoking the WMAP bispectrum up to $\omega=1000$ in a combined survey does not change this conclusion. The results are all perfectly consistent with fitting the scatter of the power spectrum estimates assuming a featureless Gaussian CMB.

It will be interesting to see how the upcoming Planck polarisation data will change these results. Oscillatory features should be more prominent in polarisation due to less severe suppression by the transfer functions so that we can expect substantially lower error bars. A scan of the Planck bispectrum should also be available in due course, allowing a combined analysis to be carried out over the entire frequency range and providing further discovery potential.

We expect that similar methods to those presented in this work can be employed to search for other feature model templates. In particular, the sharp feature limit modulations could be generalised to allow for envelopes in the power spectrum and bispectrum. Furthermore, while we focused on the bispectrum shape that arises from features in the slow-roll parameter $\epsilon$, it is also of some interest to look for the characteristic shape generated by features in the speed of sound.

\begin{acknowledgments} 
We would especially like to thank Michele Liguori who was instrumental in developing and validating the modal pipeline which was used to produce the bispectrum results presented in this paper. We would also like to thank Daniel Baumann, Xingang Chen and Yi Wang for valuable discussions and comments. We are grateful to Juha J\"{a}ykk\"{a} and James Briggs for outstanding computational support. This work is partly based on observations obtained with Planck (\url{http://www.esa.int/Planck}), an ESA science mission with instruments and contributions directly funded by ESA Member States, NASA, and Canada. HFG and BW gratefully acknowledge the support of the Studienstiftung des deutschen Volkes and STFC studentships. BW also acknowledges support from a Starting Grant of the European Research Council (ERC STG grant 279617) and from a Cambridge European Scholarship of the Cambridge Trust. This work was supported by an STFC consolidated grant ST/L000636/1. It was undertaken on the COSMOS Shared Memory system at DAMTP, University of Cambridge operated on behalf of the STFC DiRAC HPC Facility. This equipment is funded by BIS National E-infrastructure capital grant ST/J005673/1 and STFC grants ST/H008586/1, ST/K00333X/1. We acknowledge use of the HEALPix package \cite{Gorski:HEALPix}.
\end{acknowledgments}

\appendix
\section{Sharp features in single-field inflation}
\label{app:sharpfeatures}

The aim of this section is to briefly motivate the templates we use to search for sharp feature signatures in the power spectrum and bispectrum. More detailed studies of the effects of sharp features can be found elsewhere (cf.\ Refs.~\cite{Chen:PrimNonGaussianities,Dvorkin:GSR,Adshead:NonGaussianity,Bartolo:EFTfeatures} and references therein). We focus on single-field inflation with action
\begin{equation}\label{eq:action}
S=\int \mathrm{d}^4x\sqrt{-g}\left(\frac{R}{2}+P(X,\phi)\right)\,,
\end{equation}
where $P$ is an arbitrary function of $X=-1/2(\partial_{\mu}\phi)^2$ and $\phi$, $R$ is the Ricci scalar and we set the reduced Planck mass $M_{\text{pl}}=1$ for convenience.

In all of this appendix we assume that the sharp features can be dealt with perturbatively. As already discussed at the end of Sec.~\ref{sec:featuremodels} this places an upper bound on the sharpness of features that can be studied \cite{Adshead:Bounds,Cannone:PertUnitarity}. The finite width has the effect of introducing an envelope that exponentially suppresses the modulations in the power spectrum for wavenumbers that were deep inside the horizon at the time of the feature. The treatment in this appendix does not take the effects of a finite width into account and, thus, the envelopes are absent. We argued in Sec.~\ref{sec:featuremodels} that this is a reasonable idealisation when looking for the signatures of very sharp, but still perturbative features over the multipole ranges considered in this work.

\subsection{Power spectrum}
\label{subapp:sharpfeaturesPS}

From Eq.~\eqref{eq:action} one can deduce the quadratic part of the action for the scalar curvature perturbation $\zeta$
\begin{align}
S_2&=\int\mathrm{d}^4x\left(\frac{a^3\epsilon}{c_s^2}\dot{\zeta}^2-a\epsilon(\partial_i \zeta)^2\right)\\
&=\frac{1}{2}\int\mathrm{d}^3x\mathrm{d}s\left((v^{\prime})^2-(\partial_i v)^2+\frac{z^{\prime\prime}}{z}v^2\right)\,,
\end{align}
where we introduced the variable $s$ following Ref.~\cite{Hu:GSRnoncanonical} with $\mathrm{d}s=-c_s\mathrm{d}\tau=-c_s/a\,\mathrm{d}t$ and defined $v=z\zeta$ with
\begin{equation}
z^2=\frac{2\epsilon a^2}{c_s}.
\end{equation}
Intuitively, the variable $s$ measures the comoving distance sound can travel until the end of inflation at $\tau=0$. Here and in what follows $^{\prime}$ denotes derivatives with respect to $s$. Varying the action we obtain the Mukhanov-Sasaki equation of motion for the Fourier modes $v_k$:
\begin{equation}
v_k^{\prime\prime}+\left(k^2-\frac{z^{\prime\prime}}{z}\right)v_k=0\,.
\end{equation}
To lowest order in slow roll we simply have $z^{\prime\prime}/z\sim 2/s^2$. Nonetheless, this term can become very large if there is a sharp feature present at some $s_0$ in either of these parameters as it also contains the first and second derivatives of $\epsilon$ and $c_s$.

To study the behaviour of the solution for arbitrary deviations from slow roll one can make use of the GSR technique. However, since we are mainly interested in sharp features that cause high-frequency oscillations, there is an easy way to get insight into the generic behaviour of the resulting power spectra. Let us assume that $ks_0\gg1$ so that the $z^{\prime\prime}/z$-term is unimportant except in a vicinity of the sharp feature, where the derivatives can become large. We can then think of the effect of this term on such a mode as%
\footnote{Cf.\ Ref.~\cite{Bean:DualityCascade} where a similar approach was taken to derive an analytic approximation to the power spectrum modulations in the context of sharp steps in brane inflation.}%
\begin{equation}
\frac{z^{\prime\prime}}{z}\sim \frac{2}{s^2}+\frac{A}{s}\delta(s-s_0)+\frac{B}{2}\delta^{\prime}(s-s_0)
\end{equation}
for some real coefficients $A$ and $B$, where $A$ receives contributions from jump discontinuities in $\epsilon$ and $c_s$ or their derivatives. $B$ incorporates contributions proportional to $\delta^{\prime}$ and is only affected by jump discontinuities in $\epsilon$ or $c_s$, but not their derivatives. With this picture it is easy to deduce the effect of the feature. Choosing Bunch-Davies initial conditions for $s\rightarrow\infty$ we have
\begin{equation}
v(s)=\begin{cases} \frac{1}{\sqrt{2k}}\exp{(i ks)}, & s>s_0 \\
\frac{C_1}{\sqrt{2k}}\exp{(i ks)}+\frac{C_2}{\sqrt{2k}}\exp{(-i ks)}, & s<s_0 \end{cases}
\end{equation}
around $s_0$. Here, we made use of the assumption $ks_0\gg1$ so that the $z^{\prime\prime}/z$ term can be ignored except at the location of the feature and the solutions to the Mukhanov-Sasaki equation are plane waves.

To match the solutions at $s_0$ we need two boundary conditions. In order for $v_k^{\prime\prime}$ to be proportional to $\delta^{\prime}v_k$ we need a jump in $v_k$ itself,
\begin{equation}
v_k\vert_{s_0^+}-v_k\vert_{s_0^-}=\frac{B}{2}\,v_k\vert_{s_0^+}\,.
\end{equation}
This assumes that the jump is not too big so that it is justified to take $v_k\vert_{s_0^+}$ on the right-hand side of the equation. The second boundary condition can be obtained from integrating the equation across the step and taking the limit of vanishing integration range. Again assuming that the jump is small enough so that we can safely take $v_k\sim v_k\vert_{s_0^+}$ we arrive at
\begin{equation}
v^{\prime}_k\vert_{s_0^+}-v^{\prime}_k\vert_{s_0^-}=\frac{A}{s_0}v_k\vert_{s_0^+}-\frac{B}{2}\,v^{\prime}_k\vert_{s_0^+}\,.
\end{equation}
These two boundary conditions result in
\begin{align}
C_1=&1+\frac{i A}{2ks_0}\,,\\
C_2=&-\frac{1}{2}\left(B+\frac{i A}{ks_0}\right)\exp{(2i ks_0)}\,.
\end{align}
Including the correct $s\rightarrow 0$ behaviour of $v(s)$ the power spectrum is given by
\begin{equation}
P_{\mathcal{R}}(k)=\lim\limits_{ks\rightarrow 0} \left|\frac{v(s)}{z}\right|^2\sim\frac{H^2}{4k^3\epsilon c_s}\Big\vert_{ks\ll 1}\left|C_1-C_2\right|^2
\end{equation}
so that we arrive at
\begin{align}\label{eq:PSfeature}
&\quad P_{\mathcal{R}}(k)=P_{\mathcal{R},0}(k)\left(1+\Delta P_{\mathcal{R}}\right)\\\nonumber
&=P_{\mathcal{R},0}(k)\left(1-\frac{A}{ks_0}\sin{(2ks_0)}+B\cos{(2ks_0)}+\ldots\right)\,.
\end{align}

This simple calculation shows that in general single-field inflation, the dominant modulations to the power spectrum due to a sharp feature in the $ks_0\equiv \omega k\gg1$ limit are a constant cosine and a sine that is suppressed by a factor $1/(\omega k)$. Here, we identified the feature location $s_0$ with the frequency $\omega$, that we used to parametrise the oscillatory feature templates in our analysis. Equation~\eqref{eq:PSfeature} can be rewritten in the form of Eq.~\eqref{eq:modPS} by introducing the overall amplitude $A_P$ and the phase $\phi_P$.

The result is general in the sense that it applies independent of whether the feature arises due to sharp changes in the slow-roll parameter $\epsilon$ or in the speed of sound $c_s$. This is consistent with the rigorous GSR results in Ref.~\cite{Miranda:WarpFeatures}. For $\omega k\lesssim 1$ the behaviour of the solution is much more complicated and requires a rigorous GSR treatment. The S/N in the CMB at multipoles with low $l$ is poor so that nearly all the S/N for extended oscillations comes from $l>\mathcal{O}(10^2)$ corresponding to $k>\mathcal{O}(10^{-2})$. As we are mainly interested in large frequencies, $\omega\gg10^2$, this means that most of the S/N generically comes from regions with $\omega k\gg1$ so that it is justified to scan for these models in the power spectrum assuming this limit.

\subsection{Higher order correlators: the bispectrum}
\label{subapp:sharpfeaturesBS}
Higher order correlators are extracted using the in-in formalism treating the interaction terms in the higher-order actions perturbatively. In the spirit of the discussion above we only provide a brief discussion here that motivates the bispectrum shapes which are studied in this work. Thorough treatments can be found elsewhere \cite{Maldacena:NonGaussianFeatures, Chen:PrimNonGaussianities}.

The tree-level bispectrum in the in-in formalism is obtained via
\begin{align}\nonumber
&\langle \zeta_{\vec{k}_1}(t_e)\zeta_{\vec{k}_2}(t_e)\zeta_{\vec{k}_3}(t_e)\rangle\\\nonumber
:=&(2\pi)^3\delta(\vec{k}_1+\vec{k}_2+\vec{k}_3)B(k_1,k_2,k_3)\\
=&2\Re\left[-i\int\limits_{-\infty}^{t_e}\mathrm{d}t\langle\zeta_{\vec{k}_1}(t_e)\zeta_{\vec{k}_2}(t_e)\zeta_{\vec{k}_3}(t_e)H_I(t)\rangle\right]\,,
\end{align}
where $t_e$ denotes cosmic time at the end of inflation and $H_I$ is the interaction Hamiltonian. The leading-order bispectrum arises from the cubic action $S_3$. The degeneracy between sharp features in the speed of sound and features in $\epsilon$ is broken at the level of the bispectrum. In this work we specialise on features in $\epsilon$. In this case the important term is \cite{Chen:PrimNonGaussianities,Adshead:FastCompBispec}
\begin{equation}
S_3\supset \int \mathrm{d}\tau\mathrm{d}^3x \frac{1}{2}a^2\epsilon\eta^{\prime}\zeta^2 \zeta^{\prime}\,,
\end{equation}
where we simply set $c_s=1$ for simplicity so that $s=-\tau$. The leading-order bispectrum arising from the corresponding interaction Hamiltonian then is
\begin{align}\nonumber
& B(k_1,k_2,k_3)\\
=&\Re\left[i \left(\prod_i u_{k_i}(\tau_e)\right)\int\limits_{-\infty}^{\tau_e}\mathrm{d}\tau \frac{\epsilon}{\tau^2H^2}\eta^{\prime}\left(\prod_i u^{*}_{k_i}\right)^{\prime}\right]\,,
\end{align}
where $\tau_e=\tau(t_e)$ is the conformal time at the end of inflation, the $u_k(\tau)$ are the standard slow-roll mode functions
\begin{equation}
u_k(\tau)=\frac{\tau H}{\sqrt{4\epsilon k}}\left(1-\frac{i}{k\tau}\right)\exp\left(-i k\tau\right)
\end{equation}
and we made use of the fact that $a\sim-1/(\tau H)$ during inflation.

As in the previous section we investigate the case where the slow-roll parameters acquire singular behaviour so that $\eta^{\prime}$ has a $\delta$ and a $\delta^{\prime}$ component,
\begin{equation}
\label{eq:bispecsource}
\eta^{\prime}\sim C \delta(\tau-\tau_0)+D\,\tau\delta^{\prime}(\tau-\tau_0)\,.
\end{equation}
Here, $C$ and $D$ are again largely arbitrary coefficients related to the jump in the first and second derivative of $\epsilon$. The first term gives a contribution
\begin{equation}
B\sim \frac{1}{(k_1k_2k_3)^{\frac{3}{2}}}\Re\left[\frac{\epsilon}{\tau^2H^2}\left(\prod_i u^{*}_{k_i}\right)^{\prime}\Big\vert_{\tau_0}\right]\,,
\end{equation}
where we discarded factors that only affect the amplitude and do not contribute to the scale dependence.

We are interested in the behaviour for $K\tau_0\equiv(k_1+k_2+k_3)\tau_0\gg 1$ as in the case of the power spectrum. In this case the mode functions can be well approximated as
\begin{equation}
u_k(\tau)\sim\frac{\tau H}{\sqrt{4\epsilon k}}\exp\left(-i k\tau\right)
\end{equation}
and the leading-order behaviour is obtained by letting the derivative act on the exponential. This results in
\begin{align}\nonumber
B\sim&\frac{\tau_0 K}{(k_1k_2k_3)^{2}}\Re\left[i\exp\left(-i K\tau_0\right)\right]\\
\sim&\frac{(\tau_0 K)\sin(\tau_0 K)}{(k_1k_2k_3)^{2}}
\end{align}
up to an overall factor independent of $k$.

The second term in Eq.~\eqref{eq:bispecsource} containing $\delta^{\prime}$ gives
\begin{equation}
B\sim \frac{1}{(k_1k_2k_3)^{\frac{3}{2}}}\Re\left[\left(\frac{\epsilon}{\tau H^2}\left(\prod_i u^{*}_{k_i}\right)^{\prime}\right)^{\prime}\Big\vert_{\tau_0}\right]\,.
\end{equation}
Again isolating the leading-order behaviour in $\tau_0 K$ by letting the derivatives only act on the exponential in the mode functions, we obtain
\begin{align}\nonumber
B&\sim\frac{(\tau_0 K)^2}{(k_1k_2k_3)^{2}}\Re\left[\exp\left(-i K\tau_0\right)\right]\\
&\sim\frac{(\tau_0 K)^2\cos(\tau_0 K)}{(k_1k_2k_3)^{2}}\,.
\end{align}
Note that the second term also produces terms $\sim (\tau_0 K)\sin(\tau_0 K)$ similar to the first term. We can simply add these terms to the contribution arising from the first term and write the net leading-order bispectrum as
\begin{align}\nonumber
B(k_1,k_2,k_3)=&\frac{A_B}{(k_1k_2k_3)^{2}}\left(\cos{\phi_B}f_B(\omega)(\omega K)\sin(\omega K)\right.\\
&\quad\qquad\left.+\sin{\phi_B}(\omega K)^2\cos(\omega K)\right)\,,
\end{align}
where we identified the feature location with the frequency $\omega=-\tau_0$ as in the previous section. This shows that a given feature produces modulations with the same $\omega$ in both the power spectrum and the bispectrum. Here, $f_B(\omega)$ is a factor introduced to give equal S/N to the two terms and $\phi_B$ parametrises their relative contribution as in Sec.~\ref{sec:featuremodels}. 

\section{Statistics}
\label{app:statistics}

For the statistics that we apply to judge the significance of findings, the knowledge of the effective number of frequencies $N_{\text{eff}}$ in a survey or, equivalently, the effective frequency step width $\Delta\omega_{\text{eff}}$, related to $N_{\text{eff}}$ via Eq.~\eqref{eq:domeffdef}, is required. The purpose of this appendix is to determine $\Delta\omega_{\text{eff}}$ for a WMAP-like and Planck-like survey. For the combined power spectrum and bispectrum survey in Sec.~\ref{subsec:Results-combined} we also need to determine an overall $\Delta\omega_{\text{eff}}$ when combining two surveys that do not have identical $\Delta\omega_{\text{eff}}$. The latter is a straightforward extension of the methods already presented in Ref.~\cite{Fergusson:psbsfeatures}.

Based on the results in Ref.~\cite{Fergusson:psbsfeatures}, we generally assume that the effective step width is the same for a power spectrum and bispectrum survey based on the same CMB experiment (i.e.\ identical noise level and $l$-range). Hence, to determine $\Delta\omega_{\text{eff}}$ it is sufficient to focus on the computationally simpler power spectrum. Sky coverage does not have any significant effects on the correlations between models of different frequency. Hence, we create \num{10000} full sky Gaussian CMB realisations for both the WMAP and Planck scenario. In each case we use a standard concordance $\Lambda$CDM power spectrum, multiply by the beam function of the given experiment and add the appropriate noise level. In the WMAP case we use multipoles up to $l_{\text{max}}=600$ while in the Planck case we set $l_{\text{max}}=2000$, which is the $l_{\text{max}}$ used in our SMICA analysis%
\footnote{The Planck Likelihood includes multipoles up to $l=2500$ in some frequency bands, but we do not expect such high multipoles to contribute to the feature S/N due to noise and foregrounds that swamp the already heavily suppressed (due to transfer functions and lensing) signal. Hence, choosing $l_{\text{max}}=2000$ to extract $N_{\text{eff}}$ should be reliable.}. %
We then extract feature model amplitudes employing the fast quadratic estimator discussed in Sec.~\ref{subsubsec:FastQuadEst} covering frequencies up to $\omega=4000$.

\begin{figure}
 \centering
\includegraphics[width=\columnwidth]{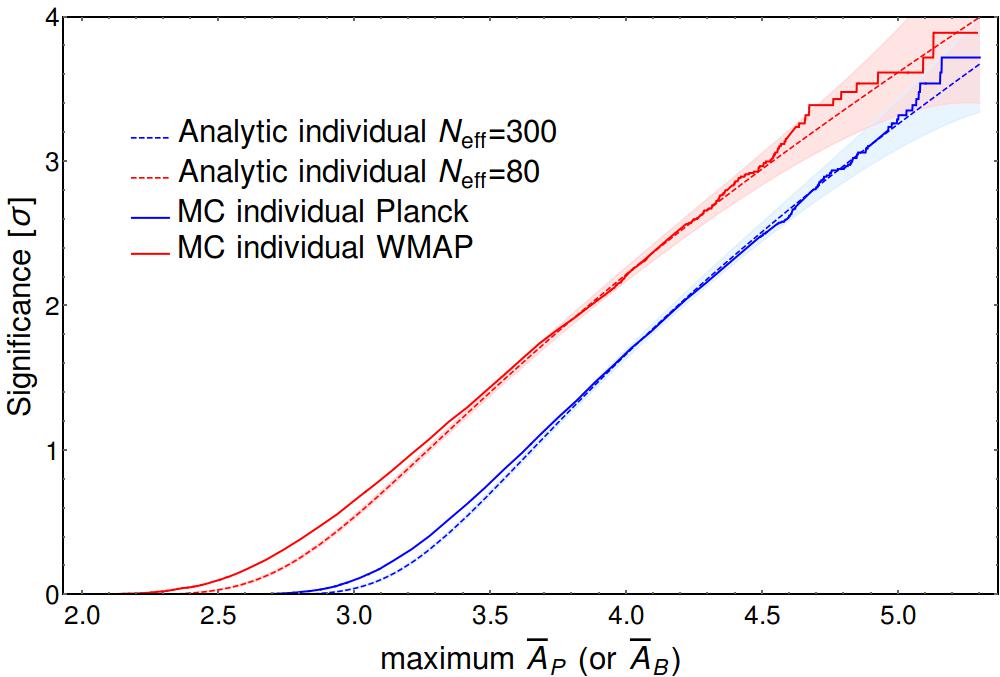}
\caption{Distribution of the maximum amplitude estimate $\bar{A}_P$ (or $\bar{A}_B$) in a power spectrum (or bispectrum) only survey covering frequencies up to $\omega=4000$ and using the templates PS1 (or BS1). We plot MC results for a WMAP-like and Planck-like survey together with the analytic models of the distributions with $N_{\text{eff}}=300$ and $N_{\text{eff}}=80$ respectively.}
\label{fig:sigindstats}
\end{figure}

Throughout this section we focus on the template PS1, i.e.\ the bare sine and cosine modulations. The corresponding distribution of the maximum significance $\bar{A}$ found in the mock surveys is shown in Fig.~\ref{fig:sigindstats}.
We also plot the analytic models of the distribution according to Eq.~\eqref{eq:sigind} for appropriate choices of the parameter $N_{\text{eff}}$. We observe very good agreement between the analytic models and the MC results as already reported in Ref.~\cite{Fergusson:psbsfeatures}. For Planck, the extracted value of $N_{\text{eff}}=300$ for this survey corresponds to $\Delta\omega_{\text{eff}}\approx 4000/300\approx 13$ while for WMAP $N_{\text{eff}}=80$ gives $\Delta\omega_{\text{eff}}\approx 50$. Note that these values are in good agreement with the general expectation that the correlation width in frequency of oscillatory modes should decay as $1/l_{\text{max}}$.

\begin{figure}
 \centering
\includegraphics[width=\columnwidth]{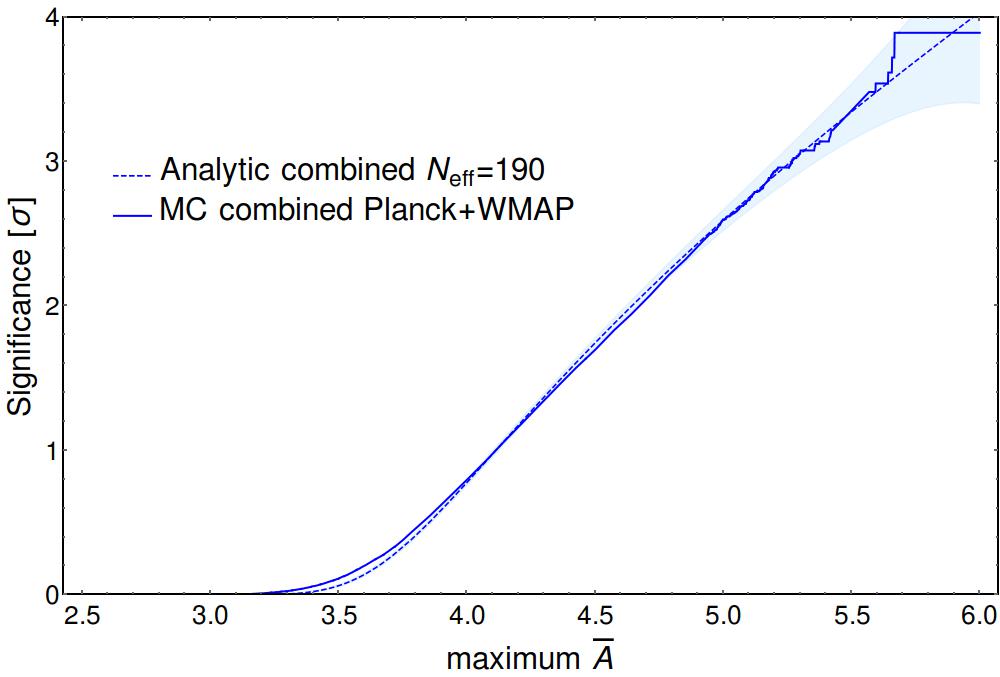}
\caption{Distribution of the maximum amplitude estimate $\bar{A}$ in a combined survey up to $\omega=4000$ where one survey is based on WMAP and the other on Planck data. We plot MC results together with an analytic model of the distribution of $\bar{A}$ in a combined survey with $N_{\text{eff}}=190$. Note that this choice of $N_{\text{eff}}$ is the mean of the corresponding values of the individual surveys.}
\label{fig:sigcombstats}
\end{figure}

To address the question of the appropriate choice of $N_{\text{eff}}$ for a combined survey we remind the reader that the statistic we are using to detect evidence for feature models is given by the maximum significance joint amplitude estimate $\bar{A}=(\bar{A}_P^2+\bar{A}_B^2)^{\frac{1}{2}}$ obtained for any relative amplitude at a given $\omega$. Just as in the case of the individual surveys we can extract an MC estimate of the distribution of the maximum $\bar{A}$ in a combined survey by using the \num{10000} WMAP-like and Planck-like realisations described above. Again, we scan for feature models up to $\omega=4000$. The corresponding results are shown in Fig.~\ref{fig:sigcombstats}.
We also plot the analytic model for the distribution of $\bar{A}$ in a combined survey, Eq.~\eqref{eq:sigcomb}, for the appropriate choice of $N_{\text{eff}}$. First of all, there is again very good agreement between MC simulations and the analytic model, showing that the model is also valid for combining surveys with different $\Delta\omega_{\text{eff}}$. This is a slight generalisation of the results presented in Ref.~\cite{Fergusson:psbsfeatures}. Furthermore, it is evident that the overall $N_{\text{eff}}$ for the combined survey can be taken to be the arithmetic mean of the values of $N_{\text{eff}}$ of the two individual surveys. In particular, in the present case we obtain $N_{\text{eff}}=(300+80)/2=190$ for the combined survey. This means that the overall $\Delta\omega_{\text{eff}}$ for a combined survey is obtained as the harmonic mean of the effective step widths of the individual surveys $\Delta\omega_{\text{eff},1}$ and $\Delta\omega_{\text{eff},2}$, i.e.\
\begin{equation}
\Delta\omega_{\text{eff}}=\frac{2}{\Delta\omega_{\text{eff},1}^{-1}+\Delta\omega_{\text{eff},2}^{-1}}\,.
\end{equation}

\bibliography{correlation.bib}

\end{document}